\documentclass[12pt]{article}
\pdfoutput=1
\usepackage{graphicx, lutypaper}
\numberwithin{equation}{section} 
\usepackage{xcolor}

\begin{document}
\begin{titlepage}

\title{Higgs Coupling Measurements\\
and the Scale of New Physics}

\author{Fayez Abu-Ajamieh}
\address{LUPM UMR5299, Universit\'e de Montpellier, 34095 Montpellier, France}

\author{Spencer Chang}

\address{Department of Physics and Institute for Fundamental Science\\ 
University of Oregon, Eugene, Oregon 97403}

\author{Miranda Chen,\ Markus A. Luty}

\address{Center for Quantum Mathematics and Physics (QMAP)\\
University of California, Davis, California 95616}

\begin{abstract}
A primary goal of present and future colliders is measuring the Higgs couplings
to Standard Model (SM) particles.
Any observed deviation from the SM predictions for these couplings is a sign
of new physics whose energy scale can be bounded from above by requiring 
tree-level unitarity.
In this paper, we extend previous work on unitarity bounds from the 
Higgs cubic coupling to  Higgs couplings to vector bosons
and top quarks.
We find that HL-LHC measurements of these couplings
compatible with current experimental bounds
may point to a scale that can be explored at the HL-LHC or a next-generation
collider.
Our approach is completely model-independent:
we assume only that there are no
light degrees of freedom below the scale of new physics, 
and allow arbitrary values for the
infinitely many couplings beyond the SM as long as they are in agreement with
current measurements.
We also extend and clarify the methodology of this analysis, and show
that if the scale of new physics is above the TeV scale, then the deviations
can be described by the leading higher-dimension gauge invariant operator,
as in the SM effective field theory.
\end{abstract}

\end{titlepage}

\noindent
\section{Introduction}
The discovery of the Higgs boson at the Large Hadron Collider (LHC) has opened a
new chapter in elementary particle physics.
For the first time, we have an experimentally established theory of particle
physics that can be consistently extrapolated to energy
scales many orders of magnitude larger than what we can hope to directly probe
experimentally.
On the other hand, there is no doubt that there is new physics beyond the
Standard Model (SM):
neutrino masses, dark matter, the matter-antimatter asymmetry, and
inflation all cannot be explained by the SM.
In addition, there are serious conceptual problems with the SM,
most importantly the absence of a natural explanation of the electroweak
scale and the cosmological constant.
Although there can be little doubt that the SM is not the 
ultimate theory of nature, none of these open questions unambiguously point 
to a scale that can be probed in future experiments.

The situation was very different before the experimental discovery of the
Higgs boson.
Unitarity arguments indicated that the theory of electroweak interactions is 
incomplete without a Higgs sector at or below the TeV scale.
It was established in the 1970s that unitarity of amplitudes at high energy 
requires the theory to be a spontaneously broken gauge theory 
\cite{LlewellynSmith:1973yud, Cornwall:1973tb, Joglekar:1973hh, Cornwall:1974km} 
(see \cite{Aoude:2019tzn, Durieux:2019eor, Bachu:2019ehv}
for a modern approach).  
Lee, Quigg, and Thacker \cite{Lee:1977yc, Lee:1977eg} turned this 
into a quantitative constraint, showing that tree-level unitarity of longitudinal vector 
boson scattering could be used to give a bound on the energy scale of
the Higgs sector
(see also \Refs{Appelquist:1987cf, Chanowitz:1985hj, Maltoni:2001dc, Dicus:2004rg}).
This bound was a major motivation 
for the Large Hadron Collider (LHC), which
indeed discovered the Higgs boson in the predicted mass range.

A very important part of the continuing high-energy collider program is the experimental study 
of the newly-discovered Higgs boson.
The Higgs boson is unlike any other elementary particle: it has spin 0 and
no other quantum numbers that distinguish it from the vacuum.
The Higgs mass has been measured at the percent level, and if the SM is assumed
to be correct, this fixes all the parameters of the theory to high accuracy.
On the other hand, the couplings of a single Higgs boson to other SM fields 
have been measured only at the $20\%$ level, while the coupling of the Higgs to itself is
only bounded to be $\lsim 10$ times the SM prediction.
Because the parameters of the SM have already been determined
much more accurately, measurements of the Higgs couplings 
are best viewed as a search for  physics beyond the SM.

All this is well known.
However, what is often not sufficiently emphasized is that if
these measurements find a deviation from the SM predictions,
then they directly point to a scale of new physics,
in exactly the same way that the work of Lee, Quigg, and Thacker pointed to the
scale of the Higgs sector before its discovery.
The reason is that the SM is the unique UV complete theory with
the observed particle content.
This means that any deviation from the SM can only be explained
by either new light degrees of freedom or new interactions that ruin the
UV completeness of the theory.
This UV incompleteness shows up in violations of tree-level unitarity, just
as for the SM without the Higgs.
Tree-level unitarity violation is a sign of strong coupling in the UV, which
requires new physics at or below that scale.

As our results will show, upcoming HL-LHC measurements of Higgs couplings probe new physics at the scale 
of a few TeV or below.  This scale is not sufficiently large that we can confidently neglect higher-dimension 
operators in the Standard Model effective field theory (SMEFT).   Therefore, in this paper we adopt a completely model-independent approach
to the interpretation of the measurements of Higgs couplings.  We describe these couplings by the following effective Lagrangian in unitary gauge:
\[
\eql{theL}
\begin{split}
\scr{L} &= \scr{L}_\text{SM}
- \de_3 \frac{m_h^2}{2v} h^3
- \de_4 \frac{m_h^2}{8v^2} h^4 
- \sum_{n \, = \, 5}^\infty \frac{c_n}{n!} \frac{m_h^2}{v^{n-2}} h^n
+ \cdots
\\
&\qquad{}
+ \de_{Z1} \frac{m_Z^2}{v} h Z^\mu Z_\mu
+ \de_{W1} \frac{2 m_W^2}{v} h W^{\mu +} W_\mu^-
+ \de_{Z2} \frac{m_Z^2}{2v^2} h^2 Z^\mu Z_\mu
+ \de_{W2} \frac{m_W^2}{v^2} h^2 W^{\mu +} W_\mu^-
\\
&\qquad\qquad{}
+ \sum_{n \, = \, 3}^\infty \left[ 
\frac{c_{Zn}}{n!} \frac{m_Z^2}{v^{n}} h^n Z^\mu Z_\mu
+ \frac{c_{Wn}}{n!} \frac{2m_W^2}{v^{n}} h^n W^{\mu +} W_\mu^-
\right] + \cdots
\\
&\qquad{}
- \de_{t1} \frac{m_t}{v} h \bar{t} t
- \sum_{n \, = 2}^\infty
\frac{c_{tn}}{n!}\frac{m_t}{v^n} h^n \bar{t}t
+ \cdots .
\end{split}
\]
Here $\scr{L}_\text{SM}$ is the SM Lagrangian,
$h$ is the real scalar field that parameterizes the physical Higgs boson
(with $\avg{h} = 0$), $Z_\mu$, $W_\mu^\pm$ are the SM gauge fields,
and $t$ is a Dirac spinor field parameterizing the top quark.
The $\de$ parameters parameterize deviations in couplings that are already
present in the SM, while the $c$ parameters denote additional couplings
that are not present in the SM.%
\footnote{The $\de$ parameters in \Eq{theL}
are directly related to the $\kappa$ parameters
used in experimental determinations of Higgs boson couplings
\cite{LHCHiggsCrossSectionWorkingGroup:2012nn}, e.g.~$\kappa_{Z}=1+\delta_{Z1}$ and $\kappa_{t}=1+\delta_{t1}$.}
The ellipses denote terms with additional derivatives and/or powers of the
SM fields.
The parameters in $\scr{L}_\text{SM}$ are measured at the percent level or
better by precision measurements of electroweak processes and the mass of the 
Higgs boson.
The parameters $\de_{V1}$ and $\de_{t1}$ are currently constrained at the
$20\%$ level, while $\de_3$, $\de_{V2}$, and $c_{t2}$ are more weakly constrained.
These couplings will be measured with significant improvements in 
accuracy at the upcoming HL-LHC run as well as at future colliders, motivating
the focus on these couplings.
As already mentioned above, any
deviation from the SM predictions in these measurements is a sign
of physics beyond the SM and points to a scale of new physics that can be
explored experimentally.
To do this, we assume that there are no additional particles below some UV
scale $E_\text{max}$, and determine $E_\text{max}$ by requiring that the
theory satisfies tree-level unitarity up to the scale $E_\text{max}$.

The implications of unitarity for extensions of the SM 
has been extensively studied, 
but there are a number of new features to the present analysis.
\begin{itemize}
\item
We use a completely model-independent bottom-up approach.
In particular, we do not make any assumption about the infinitely
many unconstrained couplings in \Eq{theL} other than that they are compatible
with existing measurements.
For example, we allow cancelations among measured and unmeasured couplings.
In this way,
we obtain unitarity constraints that are valid independently of
these parameters, and show that marginalizing over them conservatively does not
substantially improve the bounds.
\item
Previous work has focused on unitarity constraints on $2 \to 2$ partial wave amplitudes 
\cite{Lee:1977yc, Lee:1977eg, Chanowitz:1985hj,
Appelquist:1987cf, Corbett:2014ora, Corbett:2017qgl} 
and inclusive cross sections 
\cite{Chanowitz:1984ne, Maltoni:2001dc, Dicus:2004rg, Belyaev:2012bm, Falkowski:2019tft}.  
We follow \Ref{Chang:2019vez} and directly impose unitarity constraints on 
dimensionless $n \to m$ amplitudes that are generalizations of 
$2 \to 2$ partial wave amplitudes.
With this technique, we obtain unitarity bounds that can be numerically stronger
than those found in previous analyses.  In addition, these amplitudes have interesting properties, {\it e.g\/}.~potential IR enhancements and disconnected contributions, 
that merit further investigation. 
\item
We discuss the interplay between different SM deviations in determining
the scale of new physics.
For example, the dominant unitarity-violating process arising from $\de_{t1}$
also depends on $\de_{V1}$.
More phenomenologically, double Higgs production constrains a combination of
$\de_3$, $\de_{V2}$, and $c_{t2}$, and we work out the constraints on the
scale of new physics in this expanded parameter space.
\item
Without assuming
%
%
any effective Lagrangian power counting scheme,
we show that if the scale of new physics is much larger than the TeV scale, 
the deviations are well-described by the leading higher-dimension gauge 
invariant operators, as in SMEFT.
We give quantitative estimates of the errors of the SMEFT predictions
purely from unitarity.
\end{itemize}

The outline of  this paper is as follows.
In \S\ref{sec:higgsself} we consider the Higgs cubic coupling,
extending the results of \Ref{Chang:2019vez} in several ways.
First, we use this as an example to give a more detailed discussion of the 
model-independence of the unitarity bound and the effective field theory 
framework we use to obtain it.
We then show that marginalizing over unmeasured couplings does not substantially 
improve the unitarity bound, and we show that if the scale of new physics is high,
the quartic Higgs coupling is approximately described by the predictions of the
Standard Model effective field theory.
%
%
In \S\ref{sec:higgshVV} and \S\ref{sec:higgshtt} we analyze these same questions
for the $hVV$ and $h\bar{t}t$ couplings, respectively.
In these cases, we find that measurements at HL-LHC that are
consistent with current constraints may point to a scale of new physics in the
few TeV range, a scale that can be directly explored at the HL-LHC and future colliders.
In \S\ref{sec:higgsquad}, we consider the couplings $hhVV$ and $hh\bar{t}t$, 
which can also be probed by future colliders, and show that upcoming measurements
of these couplings can also point to new physics at the few TeV scale.
In \S\ref{sec:conclusions} we summarize our conclusions,
and an Appendix gives details of our calculation techniques and a summary 
of the calculations used in the main text.

\section{New Physics from the Higgs Self-Coupling}
\label{sec:higgsself}
In this section we discuss the model-independent bound on the
scale of new physics from measurements of the cubic Higgs self-coupling.
This section is based on \Ref{Chang:2019vez}, but goes beyond it in a number
of respects.
First, we include a more complete discussion of the model-independence
of the bound and the role of additional deviations from the SM that are poorly 
constrained.
Specifically, we explain why couplings with additional derivatives and powers of
gauge fields do not affect the bounds.
We also show that  marginalizing over the infinitely unmeasured
couplings does not substantially improve the bound.
Second, we show that if the scale of unitarity violation is large compared
to 1~TeV, unitarity alone implies that the deviation in the Higgs quartic
coupling is related to that of the Higgs cubic coupling as predicted by the
dimension-6 operator $(H^\dagger H)^3$.
We are able to give a quantitative estimate of the error purely from bottom-up
considerations.

\subsection{Model-Independent Bound on the Scale of New Physics}
\label{sec:hhhmodelindep}
Suppose that the experimentally measured value of the Higgs cubic coupling
differs from the prediction of the SM.
Obviously, this implies that there is physics beyond the SM, but at what 
scale?
One possibility is that this physics is near the electroweak scale, for example
additional Higgs bosons that mix with the observed Higgs boson.
In this case, the new states can be potentially produced and observed
in direct searches.
But it is also possible that the new physics responsible for the deviation is
at higher energies that are not directly probed by current experiments.
Because the SM is the unique UV complete theory with the observed particle
content, the scale of this new physics cannot be arbitrarily high.
One sign of this is that any effective theory that can explain this result
without the addition of new light particles violates tree-level unitarity
at high energies.
This scale can be computed without any additional assumptions, and gives
an upper bound on the scale of new physics.

\begin{figure}[!t]
\centerline{\begin{minipage}{0.8\textwidth}
\centering
\centerline{\includegraphics[width=350pt]{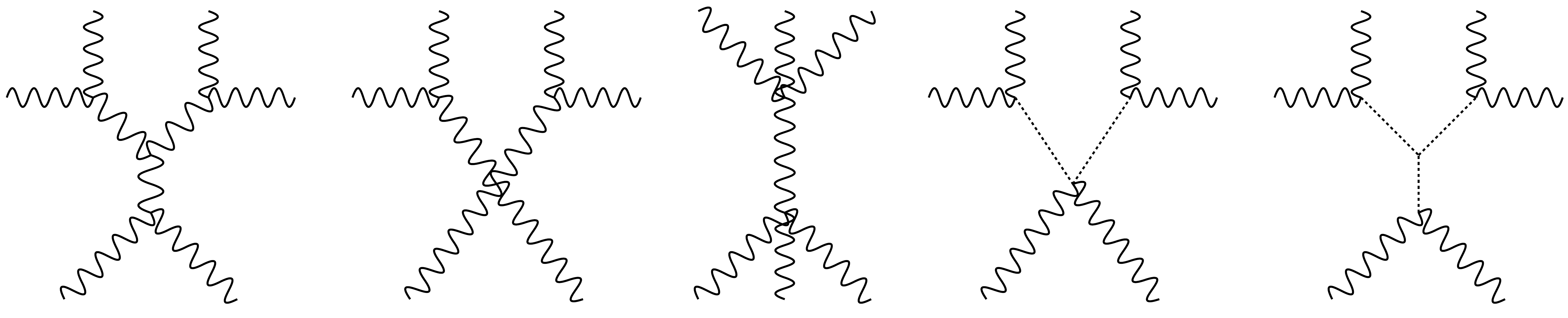}}
\caption{\small Feynman diagrams contributing to scattering processes
involving six electroweak gauge bosons.
\label{fig:sixZs}}
\end{minipage}}
\end{figure}

In a theory without gauge interactions, a cubic scalar interaction is a 
relevant coupling whose effects are small at high energies.
Nonetheless, a deviation of the Higgs cubic coupling 
from the SM prediction implies a breakdown of tree-level unitarity at 
high energies.
For example, this can be seen in the process $V_L V_L V_L  \to V_L V_L V_L $,
where $V_L $ is a longitudinally polarized $W$ or $Z$.
This has a tree-level contribution from the Higgs cubic coupling, as shown
in Fig.~\ref{fig:sixZs}.
By itself, this contributes to dimensionless amplitudes%
\footnote{We use amplitudes that are many-particle generalizations of partial
wave amplitudes normalized so that the unitarity bound is
$|\hat{\scr{M}}| < 1$.
See Appendix~\ref{sec:appendix} for details.}
with high-energy behavior $\sim E^2/v^2$, which 
would violate unitarity at high energy, but in the SM this diagram cancels with 
other diagrams to give high-energy behavior that respects unitarity.
If the Higgs cubic coupling deviates from the SM prediction, this cancellation
is destroyed, and the amplitude violates unitarity at high energies.

The scale of unitarity violation depends on the high-energy behavior of
the amplitude.
The calculation of this can be considerably simplified using the equivalence theorem,
which tells us that the leading high-energy behavior of scattering amplitudes
for longitudinally polarized gauge bosons is given by the amplitude for the 
corresponding `eaten' Nambu-Goldstone bosons \cite{Cornwall:1974km, Vayonakis:1976vz}.
We assume that experiments can be described by the effective Lagrangian 
\Eq{theL}, with no new degrees of freedom below some energy scale
$E_\text{max} \gsim \text{TeV}$.
In this section, we focus on the couplings $\de_3$ and $\de_4$ in \Eq{theL}, 
which parameterize the deviations of the  Higgs cubic 
and quartic couplings coupling from the SM values:
\[
\de_3 = \frac{g_{h^3}^{\vphantom\dagger} - g_{h^3}^\text{(SM)}}{g_{h^3}^\text{(SM)}},
\qquad
\de_4 = \frac{g_{h^4}^{\vphantom\dagger} - g_{h^4}^\text{(SM)}}{g_{h^4}^\text{(SM)}},
\]
while the $c_n$ parameters in \Eq{theL}
are couplings that are not present in the SM.

The Lagrangian \Eq{theL} is written in unitary gauge.
To use the equivalence theorem to compute the leading high-energy behavior
of amplitudes, we must restore the dependence on the Nambu-Goldstone fields.
We do this by writing the Higgs doublet in a general gauge as
\[
\eql{fullH}
H = \frac{1}{\sqrt{2}} \mat{G^1 + i G^2 \\ v + h + i G^3},
\]
where $\vec{G} = (G^1, G^2, G^3)$ parameterizes the custodial $SU(2)$ triplet of
`eaten' Nambu-Goldstone bosons.
We use a linear parameterization of the Nambu-Goldstone fields because the
SM part of the Lagrangian has manifestly good high-energy behavior
when written in terms of these fields.
To use the equivalence theorem, we must restore the dependence on the 
Nambu-Goldstone of the non-SM couplings in \Eq{theL}.
We do this by writing them in terms of the Higgs doublet \Eq{fullH}:
\[
\eql{Xdef}
X &\equiv \sqrt{2 H^\dagger H} - v
= h +  \frac{\vec{G}^2}{2(v+h)} - \frac{\vec{G}^4}{8(v+h)^3}
+ O\!\left( \frac{\vec{G}^6}{(v+h)^5} \right).
\]
Because $X = h$ in unitary gauge, the generalization of \Eq{theL} 
to a general gauge is obtained simply by the substitution $h \to X$ \cite{Chang:2019vez,Falkowski:2019tft}.
Note that $X$ is non-analytic at $H = 0$, but we are interested in the
expansion around $\avg{H} \ne 0$.

The $X^3$ term contains interactions with arbitrarily high powers of the
fields $h$ and $\vec{G}$.
However, such vertices also get contributions from  
terms of the form $X^n$ with $n \ge 4$, and these terms are 
unconstrained experimentally.
In order to obtain a bound we call our {\it model-independent bound}, we only consider processes
that do not get corrections from the unmeasured couplings $\de_n$
for $n \ge 4$.
From \Eq{Xdef} we have
\eql{Xstructure}
\[
\begin{split}
X^3 &\sim h^3 + \vec{G}^2(h^2 + h^3 + \cdots)
+ \vec{G}^4(h + h^2 + \cdots)
+ \vec{G}^6(1 + h + \cdots)
\\
&\qquad\ \gap{}
+ \vec{G}^8(1 + h + \cdots) + \vec{G}^{10}(1 + h + \cdots) + \cdots,
\\
X^4 &\sim h^4 + \vec{G}^2(h^3 + h^4 + \cdots)
+ \vec{G}^4 (h^2 + h^3 + \cdots)
+ \vec{G}^6 (h + h^2 + \cdots)
\\
&\qquad\ \gap{}
+ \vec{G}^8(1 + h + \cdots) + \vec{G}^{10}(1 + h + \cdots) +\cdots,
\\
X^5 &\sim h^5 + \vec{G}^2(h^4 + h^5 + \cdots)
+ \vec{G}^4 (h^3 + h^4 + \cdots)
+ \vec{G}^6 (h^2 + h^3 + \cdots)
\\
&\qquad\ \gap{}
+ \vec{G}^8(h + h^2 + \cdots) 
+ \vec{G}^{10}(1 + h + \cdots)
+ \cdots.
\end{split}
\]
where we have set $v=1$ and ignored numerical factors. We note that the $h \vec{G}^4$ and $\vec{G}^6$ couplings violate unitarity at high
energies, and are not affected by the unconstrained terms $X^n$
for $n \ge 4$.
We see that the unitarity-violating amplitudes that depend only on $\de_3$ are
(restoring factors of $v$)
\[
\eql{hhhform}
\hat{\scr{M}}(V_L V_L \to V_L V_L h) 
\sim \la \de_3 \frac{E}{v},
\qquad
\hat{\scr{M}}(V_L V_L V_L \to V_L V_L V_L) 
\sim \la \de_3 \frac{E^2}{v^2}.
\]
The strongest constraint comes from 
$W^+_L W^+_L W^-_L \to W^+_L W^+_L W^-_L$  and gives the bound
\[
\eql{h3bound}
E_\text{max} \simeq \frac{14\TeV}{|\de_3|^{1/2}}.
\]
For details of the calculations, see \Ref{Chang:2019vez} and the Appendix 
 of this paper.

Experimental sensitivity to a deviation in the Higgs cubic coupling comes mainly from measurements of di-Higgs production.%
\footnote{It is also possible to constrain a cubic 
deviation by looking for the $hV^4$ process in VBF production of $hV^2$ \cite{Henning:2018kys}.}
However, a deviation in this process can also be explained by new physics
contributions to the $h^2 V^2$ or $h^2 \bar{t}t$ couplings.
This will be discussed in \S\ref{sec:higgsquad} below, where we
show that a model-independent unitarity bound can be obtained by considering
these couplings together.

\subsection{Model-Independence of the Bound \label{sec:h3modelindpt}}
We claim that the bound \Eq{h3bound} is valid independently of the infinitely many
unconstrained couplings that parameterize possible deviations from the SM.
In this subsection, we discuss this point in more detail.

The discussion above has assumed that a measured deviation in the Higgs
trilinear coupling is explained by a $h^3$ coupling with no derivatives.
(The same assumption is made by the experimental searches for this deviation.)
However, there are infinitely many derivative couplings
that can contribute to an observed deviation in the Higgs cubic coupling:
\[
\De \scr{L} = \sum_{n \, = \, 1}^\infty  c_{3,n}\frac{m_h^2}{v^{2n+1}}
\d^{2n} h^3. 
\]
Here we have only shown the schematic dependence of the derivatives, but not the detailed Lorentz structure.  
If the experimentally measured $h^3$ coupling deviates from the Standard
Model prediction, this is potentially due to some 
combination of the $c_{3,n}$ couplings above.
If the deviation is dominated by a single coupling $c_{3,n}$, this requires 
\[
\frac{\de g_{h^3}}{g_{h^3}^{\text{(SM)}}}
\sim c_{3,n} \left( \frac{m_h}{v} \right)^{2n},
\]
since the Higgs coupling extraction is dominated at energies $\sim m_h$.  The $V_L^3\to V_L^3$ process leads to a unitarity violating scale (neglecting order one numerical factors)
\[
E_\text{max} \sim m_h \biggl( 
\frac{128 \pi^3 v^4}{m_h^4 } \frac{g_{h^3}^{\text{(SM)}}}{\de g_{h^3}} \biggr)^{\! 1/(2n+2)}.
\]
If one takes $\de g_{h^3} / g_{h^3}^\text{(SM)} \sim \de_3$ to compare with the earlier bound \Eq{h3bound}, one finds the unitarity bound gets more stringent with increasing $n$ and thus interpreting a Higgs trilinear deviation with the operator with the fewest derivatives leads to the most conservative new physics bound.      

An important assumption in the argument above is that the number of derivatives
in an operator determines its scaling with energy.
In particular, we assume that each additional derivative give an additional factor
of $\d \sim E$ in scattering amplitudes at high energy.
This is what is expected in general, but it can fail in certain choices
of operator basis.
This is because field redefinitions and integration by parts in the effective
Lagrangian do not affect scattering amplitudes, so there are `flat directions'
in the space of effective Lagrangians.
For example, the field redefinition
$h \to h - (\de_3/2v) h^2$
can be used to eliminate the deviation in the $h^3$ coupling, but will
induce correlated couplings of the form $h^2 \Box h, h^2 V^2$ and $h^2 \bar{t}t$.
In this basis, the $h^2\Box h, h^2V^2$ couplings typically lead to $E^4$ growth in the 
$V_L^6$ amplitude as expected from counting derivatives, but 
with the correlated values induced by the field redefinition the leading growth is canceled,
resulting in the same $E^2$ growth as the original $h^3$ deviation.
%
Thus, a basis which eliminates $h^3$ is a poor basis for our purposes, 
since it obscures the energy scaling 
through non-trivial cancellations.
To our knowledge, it has never been proven that 
there exists a basis where the \naive\ energy
scaling holds, even though this assumption is commonly used in applications
of effective field theory.
In this paper we will assume that such a basis exists, and leave further
investigation of this point for future work.%
\footnote{A natural guess is that this basis can be defined using amplitude methods \cite{Ma:2019gtx, Aoude:2019tzn}, where the connection between the number of derivatives and the energy scaling of amplitudes is manifest.}

Since the unitarity bound \Eq{h3bound} comes from scattering of gauge bosons,
we must also consider effective couplings involving gauge fields.
For example, from  the unitary-gauge diagrams shown in 
Fig.~\ref{fig:sixZs} we can see that a 
deviation  in the $hV^2$ and $h^2 V^2$ couplings can also give rise to unitarity
violation in the $V_L^6$ amplitude at high energy.
The $hV^2$ and $h^2 V^2$ couplings are phenomenologically interesting
in their own right, and will be studied in detail in \S\ref{sec:higgshVV} 
and \S\ref{sec:higgsquad} respectively below.
Here we preview some of the results of \S\ref{sec:higgshVV} to understand how 
modifications of the $hV^2$ and $h^2 V^2$ couplings contribute to
the $V_L^6$ amplitude.
To use the equivalence theorem, we restore the Nambu-Goldstone bosons
in the gauge boson fields in unitary gauge (see \Eq{VGoldstone} below):
\[
\eql{gaugeexp}
g V_\mu \to g V_\mu + \frac{\d_\mu G}{v} + \frac{h \d_\mu G}{v^2} + \cdots\,,
\]
where $g$ is the gauge coupling.
This gives (temporarily setting $v = 1$)
\[
\begin{split}\label{eq:VVXMaster}
\!\!\!\!
X (gV)^2   &\sim \partial^2 [\vec{G}^2(h + h^2+ \cdots) + \vec{G}^4(1+h  + \cdots)
+ \vec{G}^6(1 + h + \cdots) +  \cdots],
\\
\!\!\!\!
X^2 (gV)^2   &\sim \partial^2 [\vec{G}^2(h^2 + h^3+ \cdots) + \vec{G}^4(h+h^2  + \cdots)
+ \vec{G}^6(1 + h + \cdots) +  \cdots],
\\
\!\!\!\!
X^3 (gV)^2  &\sim \partial^2 [\vec{G}^2(h^3 + h^4+ \cdots) + \vec{G}^4(h^2+h^3  + \cdots)
+ \vec{G}^6(h + h^2 + \cdots) +  \cdots].
\end{split}
\]
Here we have assumed  custodial symmetry so that the Nambu-Goldstones appear in a custodial singlet $\vec{G}^2$.
These give a contribution to the $V_L^6$ amplitude
(restoring the factors of $v$)
\[
\De\hat{\scr{M}}(V_L V_L V_L \to V_L V_L V_L)
\sim (\de_{V1} + \de_{V2}) \frac{E^4}{v^4}, 
\]
where $\de_{V1}$ and $\de_{V2}$ are defined in \Eq{theL} and their coefficients in the above equation are only schematic.
We see that deviations in the $hV^2$ and $h^2 V^2$ couplings
contribute to the amplitude the same way as higher-derivative
couplings at high energy, and therefore they can only lower the scale of unitarity violation.
Similar results hold for modifications of the $V^3$ and $V^4$ couplings,
as well as terms with additional derivatives.
These give contributions to the $V_L^6$ amplitude that grow even faster with energy,
and therefore do not invalidate the bound \Eq{h3bound}.

To determine the unitarity bounds from a Higgs cubic coupling deviation, we conservatively assume that 
$\de_{V1}$, $\de_{V2}$, and higher-derivative couplings are zero and focus on the 
$\de_3$ coupling.  
Contributions to the amplitude that are higher order in $\de_3$ involve propagators that give
additional $1/E^2$ suppression at high energies, so the leading unitarity violation is given by a single
insertion of $\de_3$ even for $\de_3 \gsim 1$.%
\footnote{For other processes, we will find that the leading 
contributions to the unitarity bound include diagrams with propagators, 
for example  \Eq{tthform}.}

\subsection{The Optimal Bound\label{sec:hhhopt}}
The bound \Eq{h3bound} makes no assumption about the nature of the new physics 
other than that it is at high scales, and is valid independently of the values 
of the infinitely many unmeasured couplings $\de_4, c_n$ in \Eq{theL}.
However, it is not guaranteed this it is the best possible bound,
because it does not take the effects of all possible unmeasured couplings 
into account.
The reason is the following. If we allow additional unmeasured couplings to be nonzero, these predict
additional higher-body processes that depend on $\de_3$ as well as
the unmeasured couplings.
Requiring that these additional processes do not violate unitarity below 
the scale \Eq{h3bound} places additional constraints on these couplings.%
\footnote{In fact, we know that at least some of these couplings must
be nonzero, because the theory with only $\de_3 \ne 0$
violates unitarity at the TeV scale 
\cite{Chang:2019vez, Falkowski:2019tft}.}
It is possible that there is no choice of the new couplings that satisfies
the unitarity bound \Eq{h3bound}, in which case we obtain a stronger
unitarity bound.
In other words, an optimal bound is obtained by marginalizing over the 
unmeasured couplings, while the bound \Eq{h3bound} is independent of
these couplings.

We have not found a general method to obtain the optimal bound.
However, in the case of the $V_L^6$ amplitude we can constrain the optimal bound 
to show that it does not significantly improve the bound \Eq{h3bound}.
To do this, we consider a theory consisting of the SM plus
the dimension-6 interaction $(H^\dagger H)^3$.
This corresponds to a particular choice of the higher dimension $X^n$ 
operators that 
%
includes terms
%
%
only up to six scalars (see \Eq{fullH}).
Therefore, for this choice of couplings we can simply check all unitarity
violating processes and put a bound on the scale of unitarity violation.
The optimal bound will always be weaker than the unitarity violating scale 
obtained from the $(H^\dagger H)^3$ theory,
since this corresponds to a particular choice for the infinitely many unconstrained
couplings.
If this scale is the same as \Eq{h3bound}, we will know that this is
the optimal bound; if not, we learn that the 
optimal bound is between the bound \Eq{h3bound} and the
one just described.

\begin{figure}[!t]
\centerline{\begin{minipage}{0.8\textwidth}
\centerline{\includegraphics[width=250pt]{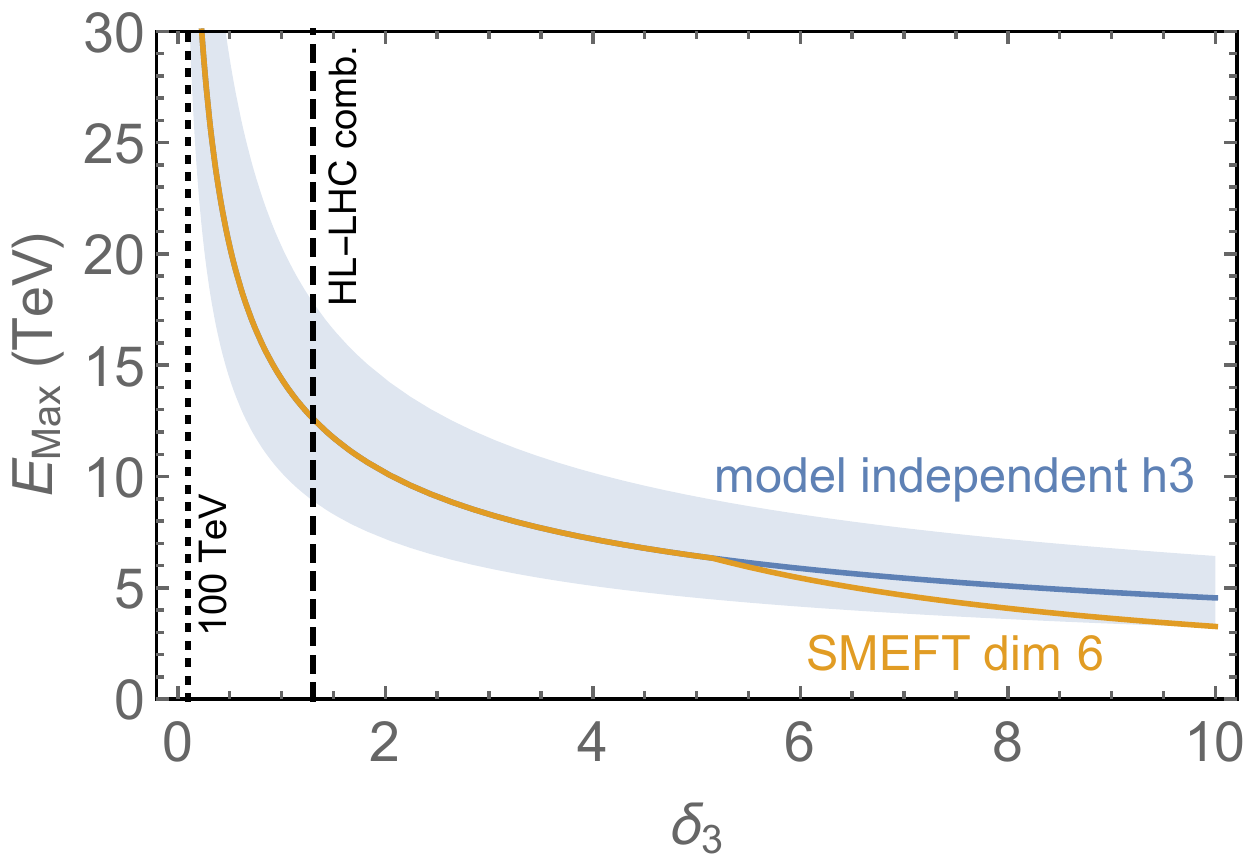}}
\caption{\small 
The unitarity bound as a function of the deviation in the $h^3$ coupling.
The optimal bound lies between the model-independent and SMEFT estimates. 
The band around the model-independent scale reflects the uncertainty of the bound from varying the unitarity
constraint to $\frac 12 \le |\hat{\scr{M}}| \le 2$.
For comparison, we show projected 95\% C.L.~limits on $\de_3$ from a combination 
at HL-LHC and a 100 TeV $pp$ collider from \cite{Cepeda:2019klc}.}
\label{fig:h3optimal}
\end{minipage}}
\end{figure}

We find that the strongest bound in the $(H^\dagger H)^3$ theory comes from
the $V_L^6$ amplitude for small values of $\de_{3}$, but for larger values the
process $hh \to hhh$ dominates and gives
\[
E_\text{max} \simeq  \frac{32\TeV}{|\de_3|}.
\]
The results are plotted in Fig.~\ref{fig:h3optimal}.
The scale of tree-level unitarity violation is an estimate for the scale of strong
coupling, and is therefore subject to theoretical uncertainty.
As a rough parameterization of this uncertainty,
we vary the constraint from $\frac{1}{2}<|\mathcal{M}| < 2$.
Within this range, we see that there is no important difference between the
model-independent bound and the optimal bound.

\subsection{SMEFT Predictions from Unitarity\label{sec:SMEFTh4}}
If the scale of new physics is high, we expect that the new physics must
be of the decoupling type.
This means that the effects of the new physics at low energies 
can be captured by adding to the SM a series of higher-dimension gauge-invariant
operators.
This is the SMEFT framework.
If experiments reveal a deviation in one or more SM measurements, without any
sign of new physics, it is most natural to interpret the results in terms
of SMEFT.

SMEFT is predictive because the same SMEFT operator controls more than
one observable.
However, these predictions assume that we can neglect higher-dimension 
terms, and the size of these corrections is unknown without further
theoretical input.
We now show that we can make an interesting quantitative statement about this 
purely from unitarity considerations.
Specifically, we show that if the scale of new physics
is much larger than the TeV scale, we can bound the error of the SMEFT prediction,
and this error bound gets better as the scale of new physics gets larger.

To be specific, we assume that $\de_3 \ne 0$, and the energy
scale of new physics is lower than some value $E_\text{max}$.
In this case, we expect that the observed deviation in the Higgs cubic
coupling can be explained by the dimension-6 SMEFT operator\footnote{Technically, this operator is a linear combination of dimension 0, 2, 4 and 6 operators, but we will refer to these linear combinations by their highest dimension.}
\[
\eql{dim6h}
\de\scr{L}_\text{SMEFT} = \frac{1}{M^2} \left(H^\dagger H-\frac{v^2}{2}\right)^3.
\]
This form of the operator  keeps the Higgs mass and electroweak VEV at their 
tree level values, but modifies the Higgs mass parameter and quartic coupling.  
If this operator dominates, it predicts
\[
\eql{SMEFThhhpred}
\de_3 = \frac{2 v^4}{M^2 m_h^2},
\qquad
\de_4 = 6 \de_3,
\qquad
c_5 = c_6 = 45\gap \de_3.
\]
We expect these predictions to become more accurate if the scale of new
physics is larger since these additional couplings themselves generate new unitarity violating amplitudes which require coupling correlations to be canceled.  

\begin{figure}[!t]
\centerline{\begin{minipage}{0.8\textwidth}
\centerline{\includegraphics[width=230pt]{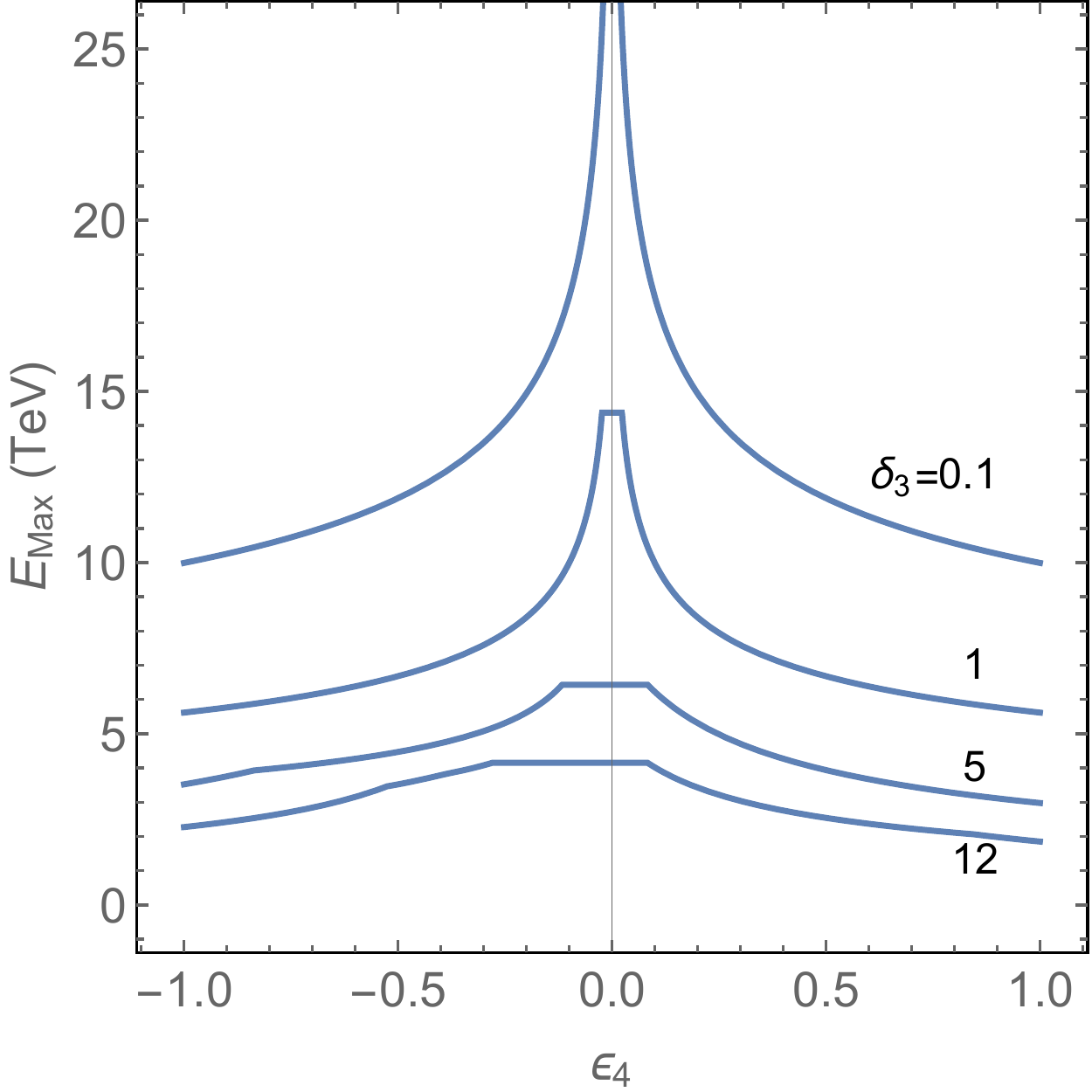}}
\caption{\small Unitarity violating scales from processes that depend on 
$\de_{3}$ and $\de_{4}$ as a function of the fractional deviation $\ep_4$
from the dimension-6 
SMEFT prediction (see \Eqs{SMEFThhhpred} and \eq{ep4}). \label{fig:epsilon4}}
\end{minipage}}
\end{figure}

To make this quantitative, we 
simply require that any deviation in the quartic coupling does not
give rise to tree-level unitarity violation below the scale $E_\text{max}$.
This requirement not only bounds the quartic coupling from being too 
large, but it also predicts that its deviation must be close to the 
prediction of the dimension-6 SMEFT operator \Eq{dim6h}:
\[
\eql{ep4}
\ep_4 = \frac{\de_4 - \de_4^{\text{dim 6}}}{\de_4^{\text{dim 6}}}
\ll 1.
\]
The reason for this is that adding a $X^4$ term to the effective Lagrangian
means that there are now additional processes that violate unitarity,
which are not affected by couplings of the form $X^n$ with $n \ge 5$.
The one that is most sensitive to new physics is the process 
$W^+_L W^+_L W^-_L W^-_L \to  W^+_L W^+_L W^-_L W^-_L$,
which gives the bound 
\[
E_\text{max} \simeq \frac{8.7\TeV}{|\de_4 - 6\de_3|^{1/4}}.
\]
The denominator vanishes for $\de_4 = 6\de_3$ because the SMEFT operator
does not contain a $\vec{G}^8$ term.
Requiring that the theory violates unitarity above some scale that is large
compared to 1~TeV therefore requires that the deviations are close to the SMEFT 
prediction $\de_4 = 6 \de_3$.
Taking into account all of the processes predicted by
the $X^3$ and $X^4$ couplings, the results are shown in 
Fig.~\ref{fig:epsilon4}.
For example, we see that for $E_\text{max} \sim 10\TeV$,
the deviation in the quartic coupling is within 
$\sim 10\%$ of the value predicted by dimension-6 SMEFT.  
This shows that not finding new physics below some scale can be 
complementary to direct searches 
\cite{Papaefstathiou:2015paa, Chen:2015gva, Belyaev:2018fky} 
in constraining the quartic coupling.

\section{New Physics from $hVV$ Couplings}
\label{sec:higgshVV}
The Higgs couplings to vector bosons $V = W^\pm, Z$  provides another sensitive probe for new physics. In this section, we work out the model-independent constraints on the scale of new physics from measurements of these couplings.
Note that we will not consider Higgs coupling to massless gauge bosons, 
which can be probed by $h\to \gamma\gamma, Z\gamma, gg$.
These lie outside the thrust of this paper because they do not lead to
high-energy growth in $V_L$ scattering.
Also, because these couplings are loop-induced in the Standard Model, we expect
that deviations from the Standard Model predictions will give rather
weak unitarity constraints.
%
%

\subsection{Model-Independent Bound on the Scale of New Physics}
It is well known that a deviation in the $hVV$ couplings leads to
unitarity violation in longitudinal $W$ and $Z$ scattering
at high energies (see \cite{Lee:1977yc, Lee:1977eg} and more 
recently~\cite{Kilian:2018bhs}).
In the SM, the Higgs exchange contribution cancels the $E^2$ growth
of other diagrams, so any modification of the $hVV$ coupling will ruin
this cancellation and lead to unitarity violation.
We can reproduce this result using the 
same model-independent bottom-up approach
we used for the $h^3$ coupling.
We write down the most general deviations from the SM involving the Higgs
and vector bosons that are quadratic in the $W$ and $Z$ gauge boson fields:
\[
\begin{split}
\scr{L} &= \scr{L}_\text{SM}
- \al \de T \left(\sfrac 12  m_Z^2 Z^\mu Z_\mu\right)
+ \de_{Z1} \frac{m_Z^2}{v} h Z^\mu Z_\mu
+ \de_{W1} \frac{2 m_W^2}{v} h W^{\mu +} W_\mu^-
\\
&\qquad{}
+ \de_{Z2} \frac{m_Z^2}{2v^2} h^2 Z^\mu Z_\mu
+ \de_{W2} \frac{m_W^2}{v^2} h^2 W^{\mu +} W_\mu^-
+ c_{Z3} \frac{m_Z^2}{3! v^3} h^3 Z^\mu Z_\mu + \cdots,
\eql{VVcoups}
\end{split}
\]
where $h$ is the scalar field that parameterizes the physical Higgs
boson (see \Eq{fullH}). As before, we do not assume any power counting for the higher terms, we only assume that their values are compatible with experimental constraints. Our bounds are obtained by marginalizing over the values of the infinitely many unmeasured couplings. For now, we do not assume that custodial symmetry is preserved by the deviations from the SM,
and therefore we have included an additional contribution to the $T$ parameter from shifting the $Z$ mass. 

To understand the implications of the couplings in \Eq{VVcoups} for processes involving longitudinally polarized vectors at high energy, we use the equivalence  theorem. To do this, we write the new couplings in \Eq{VVcoups} in terms of gauge invariant operators using
\[
\eql{Hhatdefn}
\hat{H} = \frac{H}{\sqrt{H^\dagger H}} = \mat{0 \\ 1} + O(\vec{G}).
\]
This transforms under electroweak gauge symmetry just like a Higgs doublet. This allows us to write the vector fields in terms of gauge-invariant
operators:
\[
\begin{split}
\hat{H}^\dagger i D_\mu \hat{H} &= -\frac{m_Z}{v} Z_\mu - \frac{1}{v}\d_\mu G^0
+\cdots,
\\
\tilde{\hat{H}}^\dagger i D_\mu \hat{H}
&= \frac{\sqrt{2} \gap m_W}{v} W^+_\mu + \frac{i\sqrt{2}}{v} \gap \d_\mu G^+ +
\cdots,
\\
\hat{H}^\dagger i D_\mu \tilde{\hat{H}}
&= \frac{\sqrt{2} \gap m_W}{v} W^-_\mu 
- \frac{i\sqrt{2}}{v} \gap \d_\mu G^-
+ \cdots,
\end{split}
\eql{VGoldstone}
\]
where we have defined
\[
\eql{epsHhat}
\tilde{\hat{H}} = \ep  \hat{H}^*,
\qquad
\ep = \mat{0 & 1 \\ -1 & 0}.
\]

We then use \Eq{VGoldstone} to write \Eq{VVcoups} as a sum of gauge invariant operators.
We therefore have
\[
\scr{L} = \scr{L}_\text{SM} - \frac{\al v^2 \de T}{2}
|\hat{H}^\dagger D_\mu \hat{H}|^2
+ \de_{Z1} v X |\hat{H}^\dagger D_\mu \hat{H}|^2 +  \de_{W1} v X |\tilde{\hat{H}}^\dagger D_\mu \hat{H}|^2 +\cdots,
\eql{hVVcomplete}
\]
where $X$ is defined in \Eq{Xdef}.
We can now expand this expression in powers of the Nambu-Goldstone fields $\vec{G}$
and Higgs field $h$ using
\[
\hat{H} &= 
\left(1 + \frac{\vec{G}^2}{(v+h)^2} \right)^{-1/2}
\mat{\displaystyle \frac{\sqrt{2}\gap G^+}{v + h} \\[10pt]
\displaystyle 1 + i \frac{G^0}{v + h}}
\nn
&= \mat{0 \\ 1} + \frac{1}{v+h} \mat{\sqrt{2} \gap G^+ \\ i G^0}
- \frac{\vec{G}^2}{2(v+h)^2} \mat{0 \\ 1}
+ O(\vec{G}^3).
\]
The only model-independent  couplings arising from $\de T, \de_{Z1}$
and $\de_{W1}$ are then
\[
\eql{hVVcontact}
\begin{split}
\!\!\!\!
\de\scr{L} &= \frac{\alpha \de T+\de_{Z1}}{v} h \d^\mu G^0 \d_\mu G^0
+ \frac{2\gap \de_{W1}}{v} h \d^\mu G^+ \d_\mu G^-
+ \frac{\alpha \de T}{v} (\partial_{\mu} h \partial^{\mu} G^{0})G^{0} 
\\
&\quad{}
+ \frac{i \alpha \de T}{v} \partial_{\mu}G^{0}(G^-\partial^{\mu}G^+ - G^+\partial^{\mu}G^-)
+ \frac{\alpha \de T}{2v^2}(G^+ \d_\mu G^- - G^- \d_\mu G^+)^2
\\
&\quad{}
+ \frac{2\alpha\de T+ \de_{Z1}}{2v^2} (\vec{G})^2 \d^\mu G^0 \d_\mu G^0
+ \frac{\de_{W1}}{v^2} (\vec{G})^2 \d^\mu G^+ \d_\mu G^- 
\\
&\quad{}
+ \frac{i}{v^2}  \left[(3\alpha \de T-2 \de_{W1}+2\de_{Z1}) h \d^\mu  G^0 + \alpha \de T\, G^0 \d^\mu h\right](G^+ \d_\mu G^- - G^- \d_\mu G^+) 
\\
&\quad{}
+\frac{i}{v^3}(2\alpha \de T-\de_{W1}+\de_{Z1})(\vec{G})^2 \d^\mu G^0(G^+ \d_\mu G^- - G^- \d_\mu G^+).
\end{split}
\]

Interactions involving higher powers of Nambu-Goldstone or Higgs fields can
be generated by next order couplings such as $\de_{Z2}$ and $\de_{W2}$, which are much less constrained experimentally.
Notice that the $\de T$ term contributes to these interactions at the same order as $\de_{Z1}, \de_{W1}$.  
However, given the stringent experimental constraints on the $T$ parameter, $\alpha \de T \lesssim 0.001$, these
effects are subdominant because we are considering significantly larger
deviations $\de_{Z1}, \de_{W1} \sim 0.1$, so we will often neglect $\de T$ in the following discussion.%
\footnote{\Ref{Stolarski:2020qim} recently pointed out that the $W_L W_L Z_L h$ 
amplitude violates unitarity only if custodial symmetry is broken.
This can be verified by the fourth line in \Eq{hVVcontact}.  
From the last line, we see that this also extends to the $Z_L W_L^4$ and $Z_L^3 W_L^2$ 
amplitudes.}

\begin{figure}[!t]
\centerline{\begin{minipage}{0.8\textwidth}
\centering
\centerline{\includegraphics[width=250pt]{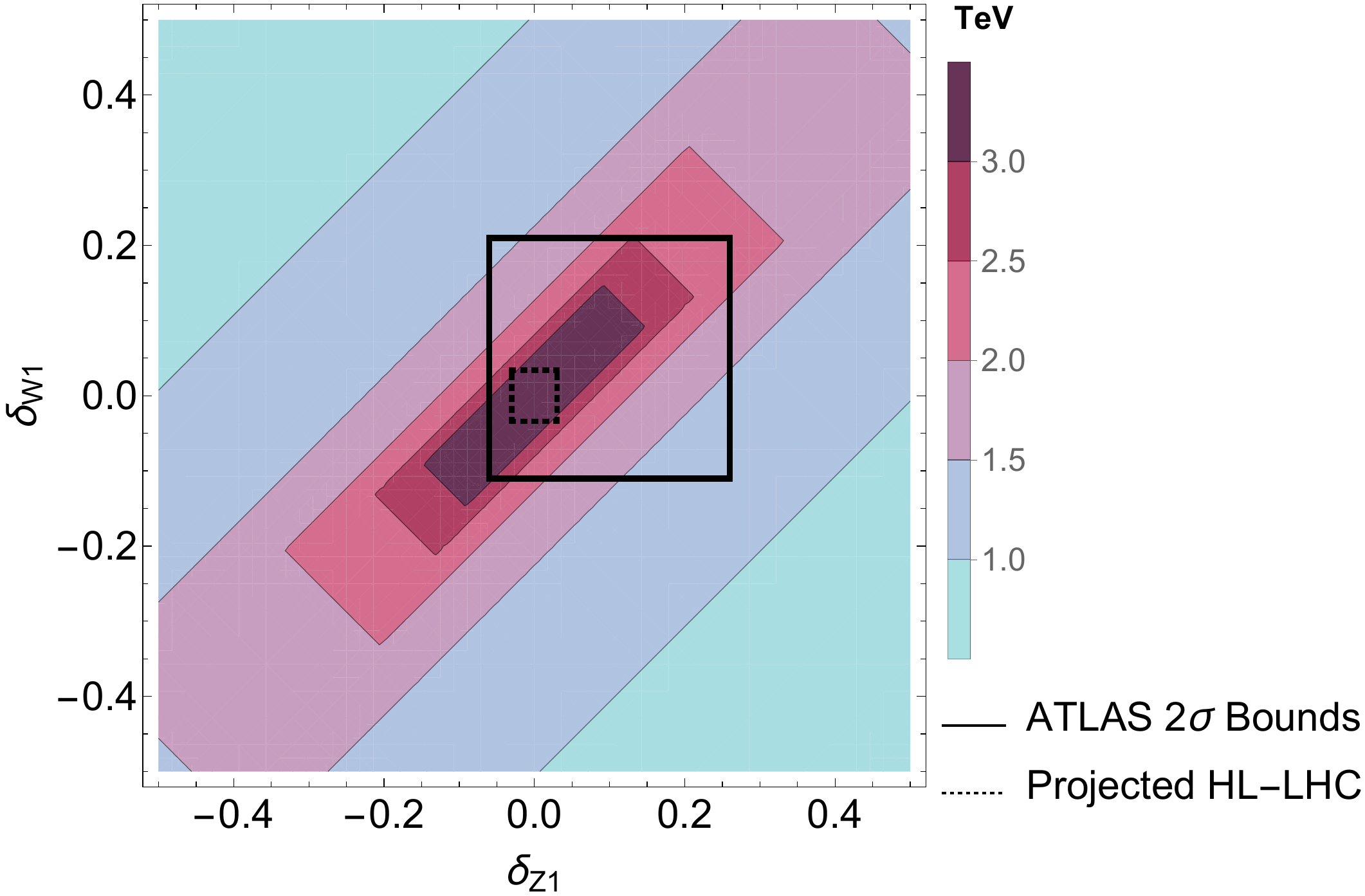}}
\caption{\small The unitarity-violating scale that depends on $\delta_{Z1}$ and $\delta_{W1}$ assuming that custodial symmetry is not preserved. The solid black line represents the current ATLAS 95\% C.L. constraints \cite{atlas:2020atl} while the dotted black line gives the HL-LHC projections \cite{Cepeda:2019klc}.}
\label{fig:CustodialViolatingPlot}
\end{minipage}}
\end{figure}

The unitarity constraints on $\de_{Z1}$ and $\de_{W1}$ come from
the amplitudes $V_L V_L \to V_L h, V_L V_L \to V_L V_L,$ and $V_L V_L V_L \to V_L V_L$.
These get contributions from a contact term from \Eq{hVVcontact} while the last two also have a contribution from 
 a Higgs exchange giving the schematic form:
\[
\begin{split}
\eql{vvhform}
\hat{\scr{M}}(V_L V_L \to V_L h) & \sim (\de_{V1})\frac{E^2}{v^2},\\ \hat{\scr{M}}(V_L V_L \to V_L V_L) & \sim (\de_{V1}+\de_{V1}^2)\frac{E^2}{v^2},\\ \hat{\scr{M}}(V_LV_LV_L\to V_LV_L) & \sim (\de_{V1} + \de_{V1}^2)\frac{E^3}{v^3}.
\end{split}
\]
Because of the experimental constraint $|\de_{V1}| \lsim 0.2$, we neglect
the quadratic terms.
The processes that give the strongest constraints are:
\[
\eql{VVVVboundsmodelindep}
\begin{split}
\!\!\!\!
W_L^+ W_L^+ \to W_L^+W_L^+
&:
E_\text{max} \simeq \frac{1.2\TeV}{|\de_{W1}|^{1/2}},
\\
Z_L Z_L \to W_L^+ W_L^- 
&: 
E_\text{max} \simeq \frac{1.5\TeV}{|\de_{Z1} + \de_{W1}|^{1/2}},
\\
W_L^+ h \to W_L^+ Z\sub{L}
&:
E_\text{max} \simeq \frac{1.0\TeV}{|\de_{Z1} - \de_{W1}|^{1/2}},
\\
W_L^+ W_L^+ W_L^- \to W_L^ + Z\sub{L}
&:
E_\text{max} \simeq \frac{1.5\TeV}{|\de_{Z1} - \de_{W1}|^{1/3}}.
\end{split}
\]
There are no unitarity constraints depending on $\de_{Z1}$ alone.
This is because the $ZZ \to ZZ$ amplitude does not grow at high energies, 
since it is proportional to $s + t + u = 4 m_Z^2$.
Note that a measured deviation on one or both of these couplings
of order of the current $2\si$ bounds $|\de_{Z1}|, |\de_{W1}| \sim 0.2$ 
would imply new physics below a few TeV,
a scale that can be explored at the HL-LHC itself. 
We plot the strongest bounds from \Eq{VVVVboundsmodelindep} in Fig.~\ref{fig:CustodialViolatingPlot}, together with the ATLAS limits on $\delta_{Z1}$ and $\delta_{W1}$ \cite{atlas:2020atl} and the HL-LHC projections \cite{Cepeda:2019klc}. Notice that $\delta_{Z1} = \delta_{W1}$ (the positive diagonal on the plot) corresponds to the custodial symmetry limit which has weaker unitarity bounds than the maximally custodial violating direction $\de_{Z1}=-\de_{W1}$, due to the last two processes in \Eq{VVVVboundsmodelindep}.

\subsection{Optimal Bound with Custodial Symmetry}\label{sec:custodialPres}
As emphasized in \S\ref{sec:hhhmodelindep},
bounds such as \Eq{VVVVboundsmodelindep} make
no assumptions about the nature of the new physics other than that it is at
high scales, and are valid independently of the values of the infinitely 
many unmeasured couplings.
However, as discussed in \S\ref{sec:hhhopt},
marginalizing over these unmeasured couplings may give a stronger
bound, which we call the optimal bound.
In this section we show that if we assume that the new physics preserves
custodial symmetry, the model-independent bound from 
\Eq{VVVVboundsmodelindep} is in fact optimal.
We will discuss the case without custodial symmetry in 
\S\ref{sec:hVVcustviol} below.

\begin{figure}[!ht]
\centerline{\begin{minipage}{0.8\textwidth}
\centering
\centerline{\includegraphics[width=250pt]{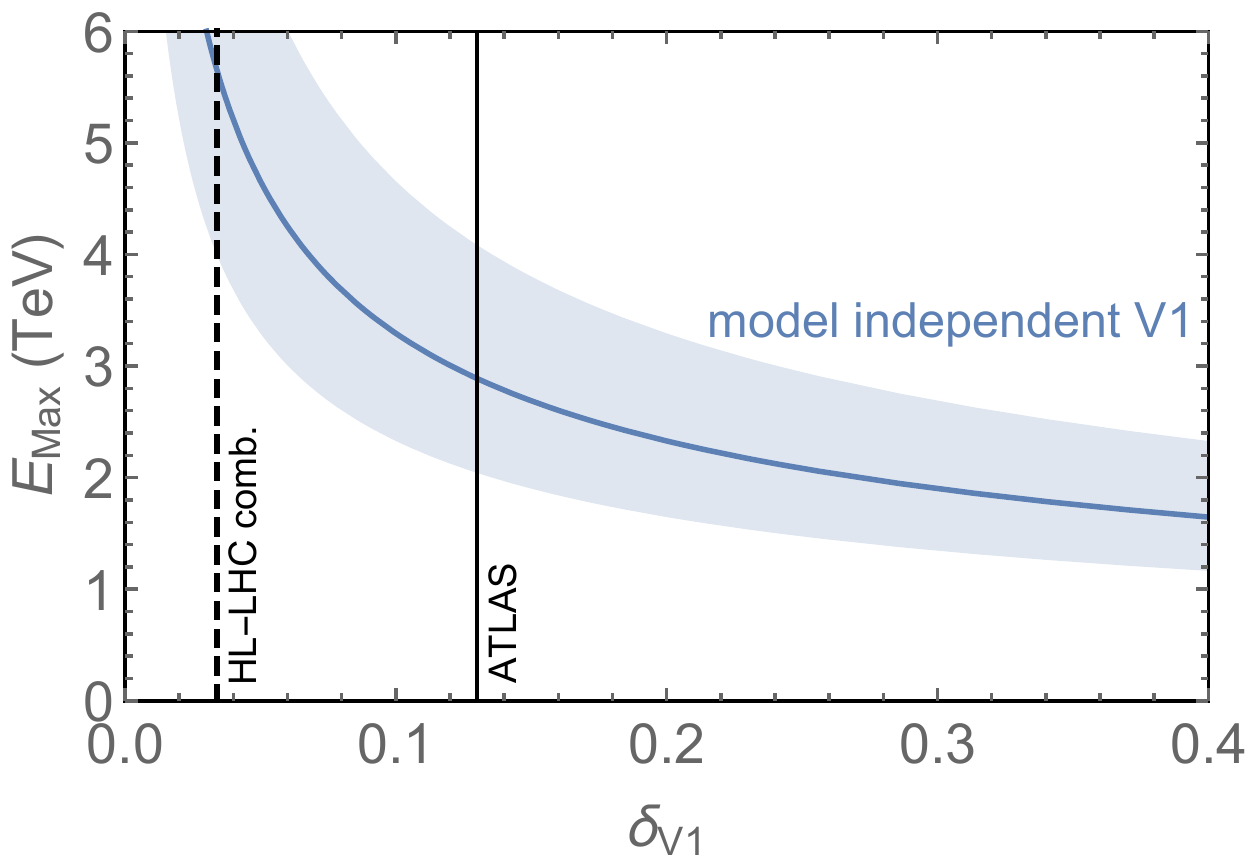}}
\caption{\small The unitarity bound as a function of the deviation in the $hVV$ 
coupling.  The optimal bound lies between the model-independent and SMEFT estimate
from the dimension-6 operator \Eq{VVhSMEFT} and thus they are the same.  
The band around the model-independent scale results from varying the unitarity
bound to $\frac 12 \le |\hat{\scr{M}}| \le 2$.
For comparison, we show the 95\% C.L.~limits on $\de_{V1}$ from 
ATLAS \cite{atlas:2020atl} and a projected HL-LHC combination \cite{Cepeda:2019klc}.
\label{fig:SMEFTV1}}
\end{minipage}}
\end{figure}

We focus on the custodial symmetry limit where
$\de T = 0$ and  $\de_{W1} = \de_{Z1} \equiv \de_{V1}$.
This limit is well-motivated by the strong experimental bounds on the $T$
parameter.
We consider the dimension-6 SMEFT operator
\[
\eql{VVhSMEFT}
\de\scr{L}_\text{SMEFT} =
\frac{1}{M^2}\left(H^\dagger H - \frac{v^2}{2}\right)|D_{\mu}H|^2.
\]
This does not contribute to the $T$ parameter, and gives a
custodial symmetry preserving deviation to the $h VV$ couplings. 
Making a field redefinition to remove the momentum-dependent terms
$h\partial h^2$ and $h^2 \partial h^2$, we find that this operator predicts
\[
\eql{VVhpredic}
\de_{V1}= \frac{v^2}{2M^2},\quad \de_{V2}=4\de_{V1}, \quad c_{V3} = 8\de_{V1}, \quad c_{V4} = 8\de_{V1},
\]
where $\de_{V2} = \de_{Z2} = \de_{W2}$,
and $c_{Vn} = 0$ for $n \ge 5$.
Using this, we can calculate the additional amplitudes predicted by \Eq{VVhSMEFT}
that violate unitarity, namely $h^2 Z_L^2$ and $h^2W_L^2$ and check whether these
give a lower scale of unitarity violation for a given value of $\de_{V1}$.
We find that these new processes give weaker or equivalent bounds to the model-independent
bound for $\de_{Z1} = \de_{W1}$, 
\[
E_\text{max} \simeq \frac{1.1\TeV}{|\de_{V1}|^{1/2}},
\]
which is therefore also the optimal bound in this case.  This is shown in Fig.~\ref{fig:SMEFTV1} along with the constraints from ATLAS and a HL-LHC projection, showing the potential to constrain new physics below $\sim 5$ TeV.

\subsection{SMEFT Predictions from Unitarity with Custodial Symmetry
\label{sec:SMEFTV2}}
\begin{figure}[!t]
\centerline{\begin{minipage}{0.8\textwidth}
\centering
\centerline{\includegraphics[width=250pt]{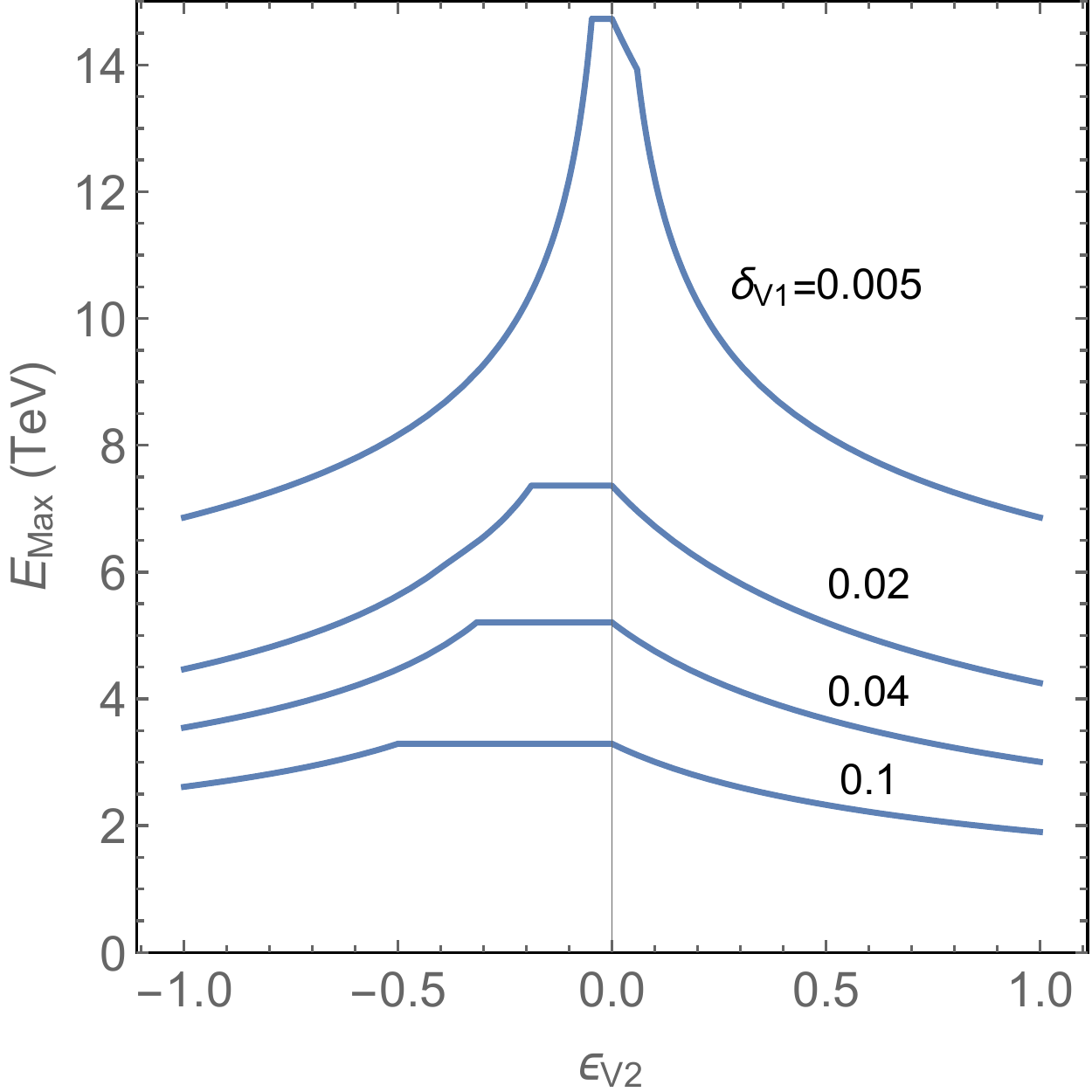}}
\caption{\small Unitarity violating scales from processes that depend on $\de_{V1}$ and $\de_{V2}$ as a function of the fractional deviation of $\de_{V2}$ from its SMEFT prediction, $\de_{V2} = 4\de_{V1}(1+\epsilon_{V2})$. \label{fig:v1epsilon2}}
\end{minipage}}
\end{figure}

If the scale of new physics is high, we expect that an observed deviation
in the Higgs couplings can be described by the lowest-dimension SMEFT operator.
In this section we assume that the new physics preserves custodial symmetry,
and consider the question of the accuracy of the SMEFT prediction, 
following the logic explained in \S\ref{sec:SMEFTh4}.
The dimension-6 SMEFT operator \Eq{VVhSMEFT} predicts $\de_{V2} = 4\de_{V1}$, 
and we define
\[
\eql{hVVdeviation}
\epsilon_{V2} \equiv \frac{\de_{V2}-\de_{V2}^{\text{dim 6}}}
{\de_{V2}^{\text{dim 6}}}.
\]
When we include both $\de_{V1}$ and $\de_{V2}$, we have the additional 
model-independent processes
$hh\to V_L V_L$,  $hV_L V_L  \rightarrow V_L V_L$ and 
$V_L V_L V_L \rightarrow V_L V_L V_L$.
Requiring that these do not violate unitarity
constrains $E_\text{max}$ for a given value of $\ep_{V2}$.
The results are shown in Fig.~\ref{fig:v1epsilon2}. 
The results are qualitatively similar to the case of the Higgs self-interaction.
The predictions of SMEFT become accurate for $E_\text{max} \gsim 10\TeV$,
corresponding to values of $\de_{V1}$ much smaller than what will be probed in upcoming experiments, and since the unitarity-violating scale is low even for  $\delta_{V1}$ of $O$(1\%), in this case a general value of $\delta_{V2}$ does not change the bound much.

\subsection{Optimal Bound Without Custodial Symmetry}
\label{sec:hVVcustviol}
We now consider the unitarity bounds for the case  $\de_{Z1} \neq \de_{W1}$.  This case is somewhat unnatural, in the sense that for values of $\de_{Z1}$ and $\de_{W1}$ that violate custodial symmetry at a level that is  observable in upcoming experiments, the small observed $T$ parameter appears to require an unnatural cancellation. 
Nonetheless, $\de_{Z1}$ and $\de_{W1}$ will be independently measured, and it is
interesting to explore the implications of $\de_{Z1} \neq \de_{W1}$.

\begin{figure}[!ht]
\centerline{\begin{minipage}{0.8\textwidth}
\centering
\centerline{\includegraphics[width=250pt]{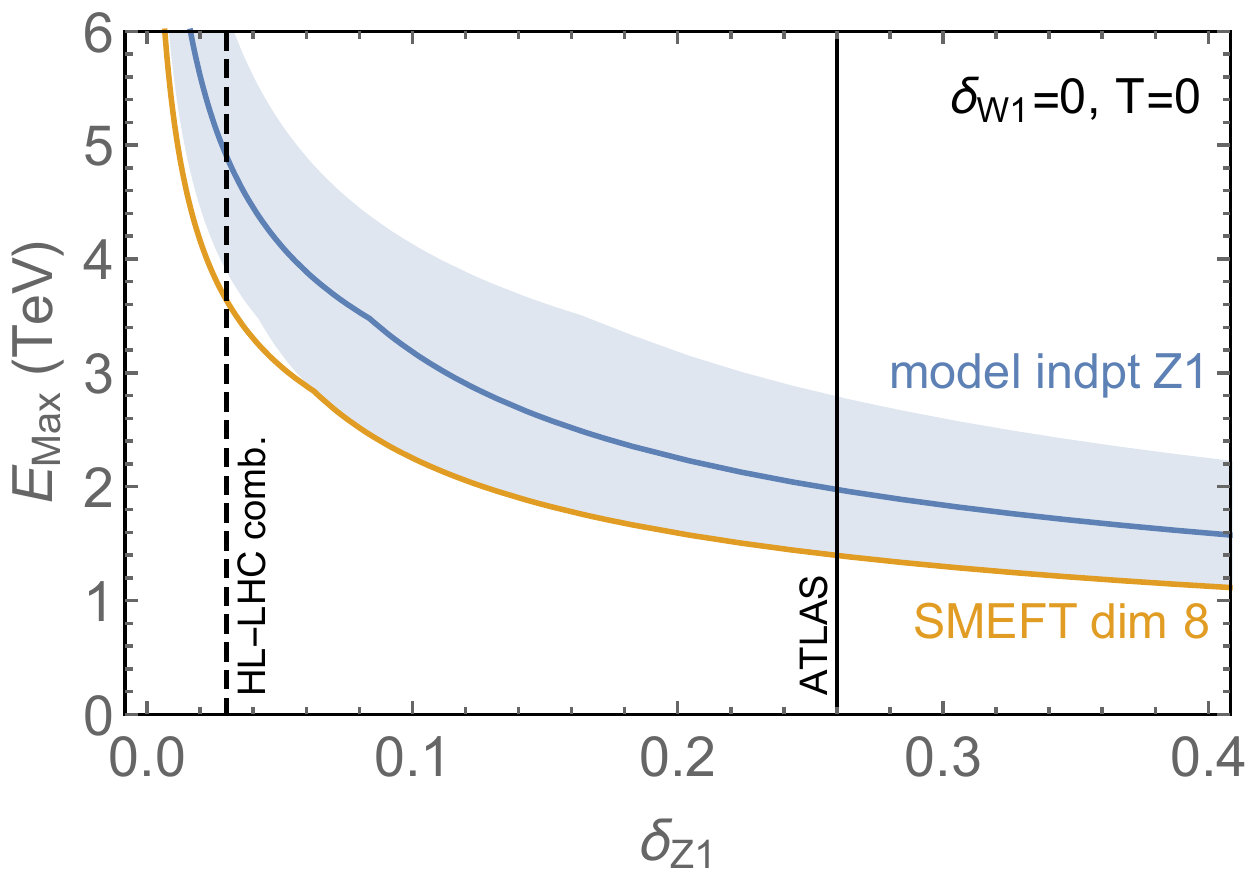}}
\caption{\small The unitarity bound as a function of the deviation in the $hZZ$ 
coupling, assuming $\de_{W1} = 0$, $\de T=0$.
The optimal bound lies between the model-independent and SMEFT estimate
from the dimension-8 operator \Eq{ZZhSMEFT}.
The band around the model-independent scale results from varying the unitarity
bound to $\frac 12 \le |\hat{\scr{M}}| \le 2$.
For comparison, we show the 95\% C.L.~limits on $\de_{Z1}$ from 
ATLAS \cite{atlas:2020atl} and a projected HL-LHC combination \cite{Cepeda:2019klc}.
\label{fig:SMEFTZ1}}
\end{minipage}}
\end{figure}

For concreteness we consider the case 
$\de_{Z1} \neq 0$, $\de_{W1} \simeq 0$, $\alpha \de T \simeq 0$.  
In order to explain this in SMEFT, we must introduce the dimension-8 operator
\[
\eql{ZZhSMEFT}
\frac{1}{M^4}\left(H^\dagger H - \frac{v^2}{2}\right)|H^\dag D_{\mu}H|^2,
\]
which has been chosen so that $\de T=0$.
This operator predicts the following coupling deviations:
\[
\eql{ZZhpredic}
\de_{Z1} &=\frac{v^4}{4M^4},
& \de_{W1} &= 0,
& \de_{Z2} &= 8 \de_{Z1},
& \de_{W2} &= -\de_{Z1},
\nn
c_{Z3} &= 40 \de_{Z1},
& c_{W3} &= -8 \de_{Z1},
& c_{Z4} &= 136 \de_{Z1},
& c_{W4} &= -32 \de_{Z1},
\\
c_{Z5} &= 288 \de_{Z1},
& c_{W5} &= -72\de_{Z1},
& c_{Z6} &= 288\de_{Z1}, 
& c_{W6} &= -72\de_{Z1}.
\nonumber
\]
There are now many more unitarity-violating amplitudes, and the unitarity violating
scale that we obtain assuming that the dimension-8 operator dominates is
somewhat stronger than the model-independent bound.
The results are shown in Fig.~\ref{fig:SMEFTZ1}.

\section{New Physics from $h\bar{t}t$ Couplings}
\label{sec:higgshtt}
The Higgs couplings to top quarks $h\bar{t}t$ provides another sensitive probe of 
new physics.
In this section we work out the model-independent constraints on the scale of new
physics from measurements of this coupling.

\subsection{Model-Independent Bound}
\begin{figure}[!t]
\centerline{\begin{minipage}{0.8\textwidth}
\centering
\centerline{\includegraphics[width=220pt]{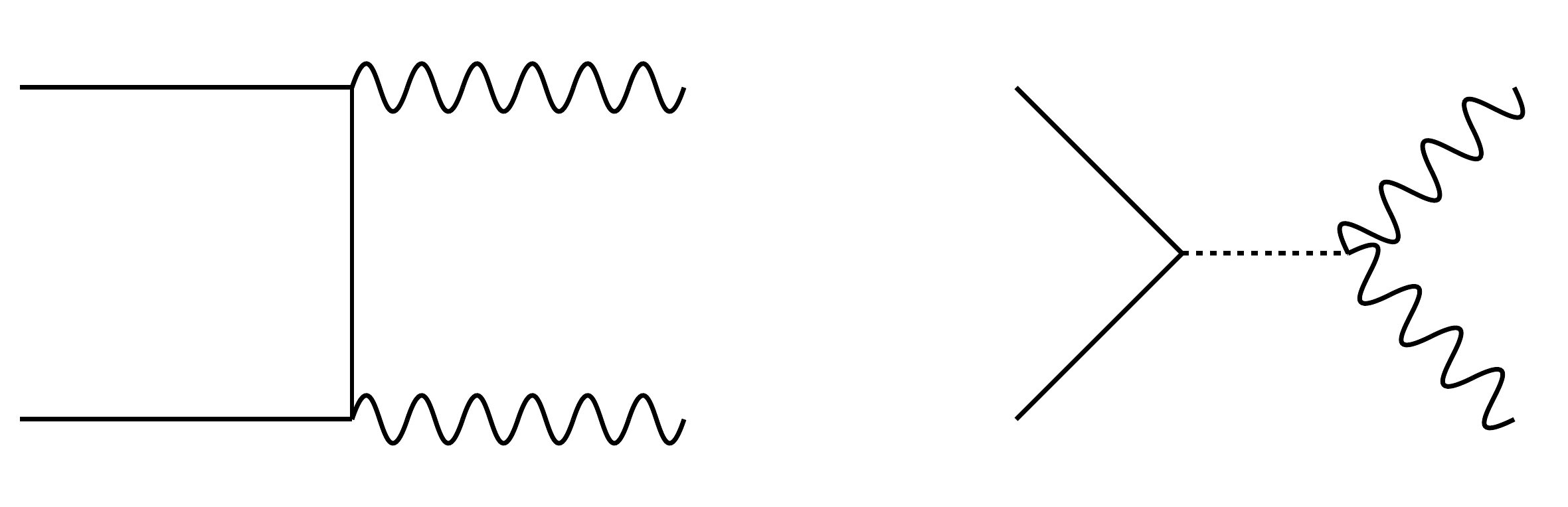}}
\caption{\small Feynman diagrams contributing to $t \bar{t} \to W^+_L W^-_L$ in unitary gauge.
\label{fig:ttWW}}
\end{minipage}}
\end{figure}

If the $h\bar{t}t$ coupling deviates from the SM value, processes such as
$t \bar{t} \to W^+_L W^-_L$ will violate unitarity at high energy.
This observation goes back to \Ref{Appelquist:1987cf}, 
which put a bound on the scale of fermion mass generation in a theory 
without a Higgs boson.
The diagrams contributing to this process in unitary gauge are shown in 
Fig.~\ref{fig:ttWW}.
We see that they are sensitive to both the $\bar{t}t h$ coupling and the
$hVV$ coupling, and we will see that the unitarity bound depends
on both $\de_{t1}$ and $\de_{V1}$ in \Eq{theL}.
Unitarity violation for more general top couplings in $2 \to 2$ processes has
been recently studied in \cite{Dror:2015nkp, Maltoni:2019aot}.

As in the previous sections, we use the equivalence theorem to compute the
high-energy behavior of amplitudes involving longitudinally polarized
vector bosons and Higgs fields.
We do this by writing the deviations from the SM in \Eq{theL} that depend
on the top quark in a general gauge:
\[
\eql{httcomplete}
\de\scr{L} = - m_t 
(\bar{Q}_L \tilde{\hat{H}} t^{\vphantom\dagger}_R + \hc)  \left(\de_{t1} \frac{X}{v}  +c_{t2} \frac{X^2}{2! v^2} +\cdots \right),
\]
where $X$ is given by \Eq{Xdef}
and $\tilde{\hat{H}}$ is given by \Eqs{Hhatdefn} and \eq{epsHhat}.
Expanding these terms in terms of the Higgs and Nambu-Goldstone bosons gives
\[
\bar{Q}_L \tilde{\hat{H}} t^{\vphantom\dagger}_R + \hc 
&= \frac{1}{\sqrt{1 + \frac{\vec{G}^2}{(v+h)^2}}}
\left(\bar{t}t - \frac{1}{v + h}\left[G^0  \bar{t} i\ga_5 t + \sqrt{2} G^-  \bar{b}_L t_R + \sqrt{2} G^+  \bar{t}_R b_L \right] \right).
\]
This leads to the following interaction pattern (temporarily setting $v = 1$)
\[
\begin{split}
\!\!\!\!\!
\bar{t}t X &\sim tt^c [h+ iG^0 (h+ \cdots) + \vec{G}^2(1+  \cdots) + i G^0 \vec{G}^2(1+ \cdots)
+ \vec{G}^4(1 +   \cdots)  
+  \cdots] 
\\
&\qquad{}
+ bt^c G^+ [(h + \cdots)+ \vec{G}^2(1  + \cdots) 
+ \vec{G}^4(1   + \cdots) + \cdots ] +\hc,
\\
\!\!\!\!\!
\bar{t}t X^2  &\sim tt^c [h^2+ iG^0 (h^2+ \cdots) + \vec{G}^2(h+  \cdots) + i G^0 \vec{G}^2(h+ \cdots)
+ \vec{G}^4(1 +   \cdots)  
+  \cdots]
\\ 
&\qquad{}
+ bt^c G^+ [(h^2 + \cdots)+ \vec{G}^2(h  + \cdots) 
+ \vec{G}^4(1   + \cdots) + \cdots ] +\hc,
\\
\!\!\!\!\!
\bar{t}t X^3  &\sim tt^c [h^3+ iG^0 (h^3+ \cdots) + \vec{G}^2(h^2+  \cdots) + i G^0 \vec{G}^2(h^2+ \cdots)
+ \vec{G}^4(h +   \cdots)  
+  \cdots] 
\\
&\qquad{}
+ bt^c G^+ [(h^3 + \cdots)+ \vec{G}^2(h^2  + \cdots) 
+ \vec{G}^4(h   + \cdots) + \cdots ] +\hc,
\eql{ttstructure}
\end{split}
\]
where the parentheses allow arbitrary higher powers of $h$.  
Examining the structure of the interactions in \Eq{ttstructure}, 
we see that the model-independent couplings that depend only on $\de_{t1}$ are
\[
\eql{httmodelindep}
\de\scr{L}
\supset& -\de_{t1} \frac{m_t}{v} \left[ 
\left(h+ \frac{1}{2v}\vec{G}^2\right)  \ggap \bar{t}t 
-\left(h +\frac{1}{2v}  \vec{G}^2  \right) \frac{G^0}{v}  \gap \bar{t} i\ga_5 t\right] \\
& + \de_{t1} \frac{\sqrt{2}m_t}{v^2} \left[\left(h  + \frac{1}{2v} \vec{G}^2\right) G^- \bar{b}_L t_R+\hc \right].
 \nonumber
\]
As discussed previously in \S\ref{sec:h3modelindpt}, we can also consider $tth$ interactions with additional derivatives,
but again we expect these will give a parametrically lower scale of unitarity violation,
and therefore in terms of new physics bounds, it is conservative to interpret a $tth$ coupling deviation in terms of the coupling with no derivatives.
We can then determine the schematic form for the following model-independent 
amplitudes:
\[
\eql{tthform}
\begin{split}
\hat{\scr{M}}(\bar{q} q \to V_LV_L) &\sim y_t\left(\de_{t1}+\de_{V1}+\de_{t1}\de_{V1}\right)\frac{E}{v},
\\
\hat{\scr{M}}(\bar{q} q \to V_L h ) &\sim y_t \left(\de_{t1}+ \de_{V1}\right)\frac{E}{v},
\\
\hat{\scr{M}}(\bar{q} q \to V_L V_L V_L) &\sim y_t\left(\de_{t1}+\de_{V1}+\de_{t1}\de_{V1}+\de_{V1}^2\right)\frac{E^2}{v^2},
\end{split}
\]
where $q = t, b$.
For the $\bar{b}t$ initial state processes, the first process vanishes.
Amplitudes related to these by crossing have the same scaling.
The terms depending on $\de_{V1}$ arise from diagrams with propagators
(see \Eq{hVVcontact}).
The 2 derivatives in vertices from $\de_{V1}$
cancel the energy suppression of the extra propagators, so these contributions
are the same order.
For contributions with a propagator, there is a possibility of $\log(E/m)$
terms arising from the phase space integrals in the amplitudes.
By direct calculation,  we show that these are absent in all of the terms in \Eq{tthform},
except possibly for the $\de_{V1}^2$ term in the last line.
This contribution is numerically small even if a log is present,
and so we will neglect all quadratic contributions.

\begin{figure}[!t]
\centerline{\begin{minipage}{0.8\textwidth}
\centering
\centerline{\includegraphics[width=300pt]{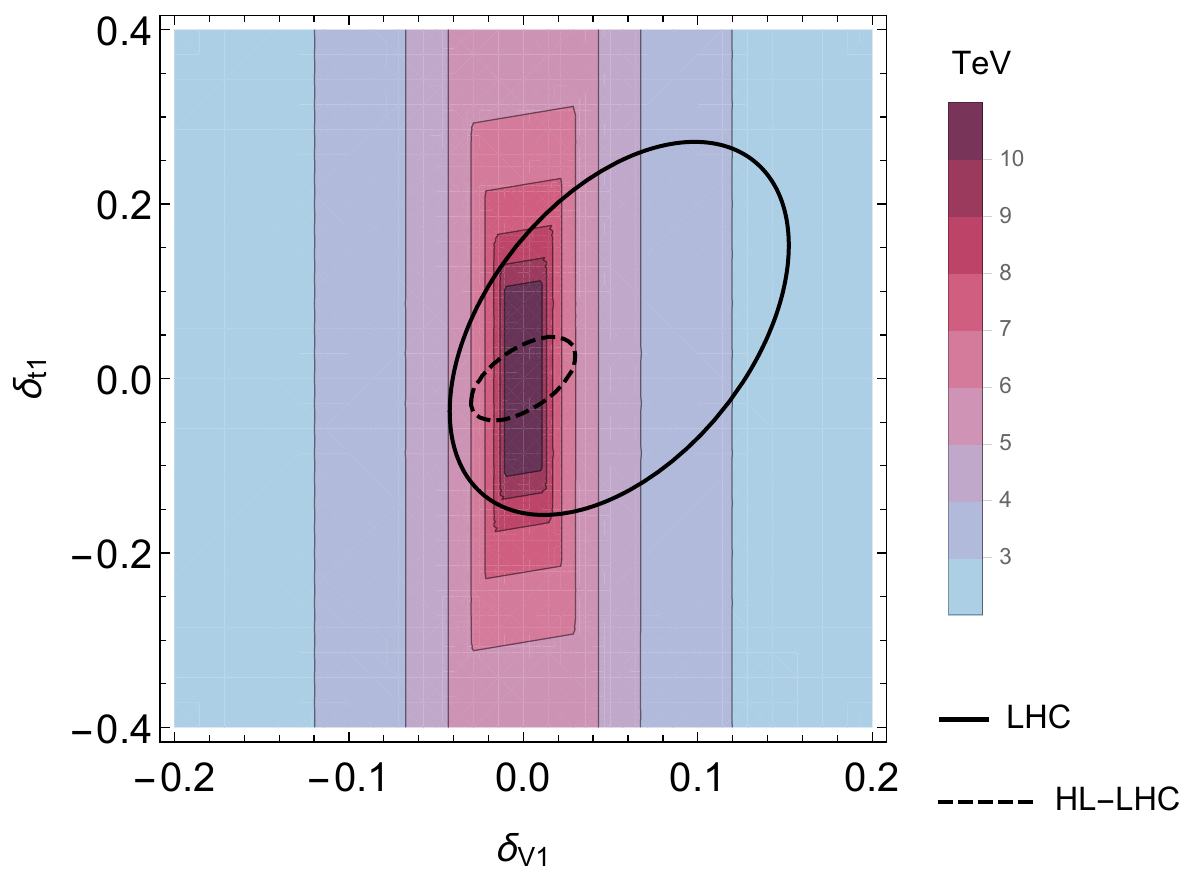}}
\caption{\small Unitarity violating scales given values of $\de_{t1}$ and $\de_{V1}$. The solid line represents the 95\% C.L.~at the LHC \cite{atlas:2020atl} and the dashed line is the HL-LHC projection for ATLAS \cite{atlas:2018prj}. \label{fig:t1andv}}
\end{minipage}}
\end{figure}

The best bounds on $\de_{t1}$ from these processes are
\eql{t1modelindpt}
\[
\begin{split}
t_R \bar{t}_R \to W^+_L W^-_L &:
E_\text{max} \simeq \frac{5.1\TeV}{|\de_{t1} + \de_{V1}|},
\\[5pt]
t_R \bar{b}_R \to W^+_L h &: E_\text{max}  
\simeq \frac{3.6 \TeV}{|\de_{t1}-\de_{V1}|}, 
\\[5pt]
t_R \bar{b}_R \to W^+_L W^+_L W^-_L &: E_\text{max}  
\simeq \frac{3.3 \TeV}{\sqrt{|\de_{t1}-\frac{1}{3}\de_{V1}|}},
\end{split}
\]
where we assume custodial symmetry $\de_{Z1} = \de_{W1} = \de_{V1}$.
As already mentioned above, these bounds are numerically stronger than
previous bounds \cite{Appelquist:1987cf,Maltoni:2001dc, Dicus:2004rg}.

Fig.~\ref{fig:t1andv} shows the unitarity violating scale from these processes
as a function of $\de_{t1}$ and $\de_{V1}$, together with projected HL-LHC
constraints on these couplings.
From this graph, we see that upcoming measurements of $\de_{V1}$ are sensitive
to lower scales of new physics.
However, if measurements of $hVV$ agree with the SM, a deviation in the
$h\bar{t}t$ coupling at HL-LHC that is compatible with current constraints
can still point to a scale of new physics below 8~TeV.

\begin{figure}[!t]
\centerline{\begin{minipage}{0.8\textwidth}
\centering
\centerline{\includegraphics[width=250pt]{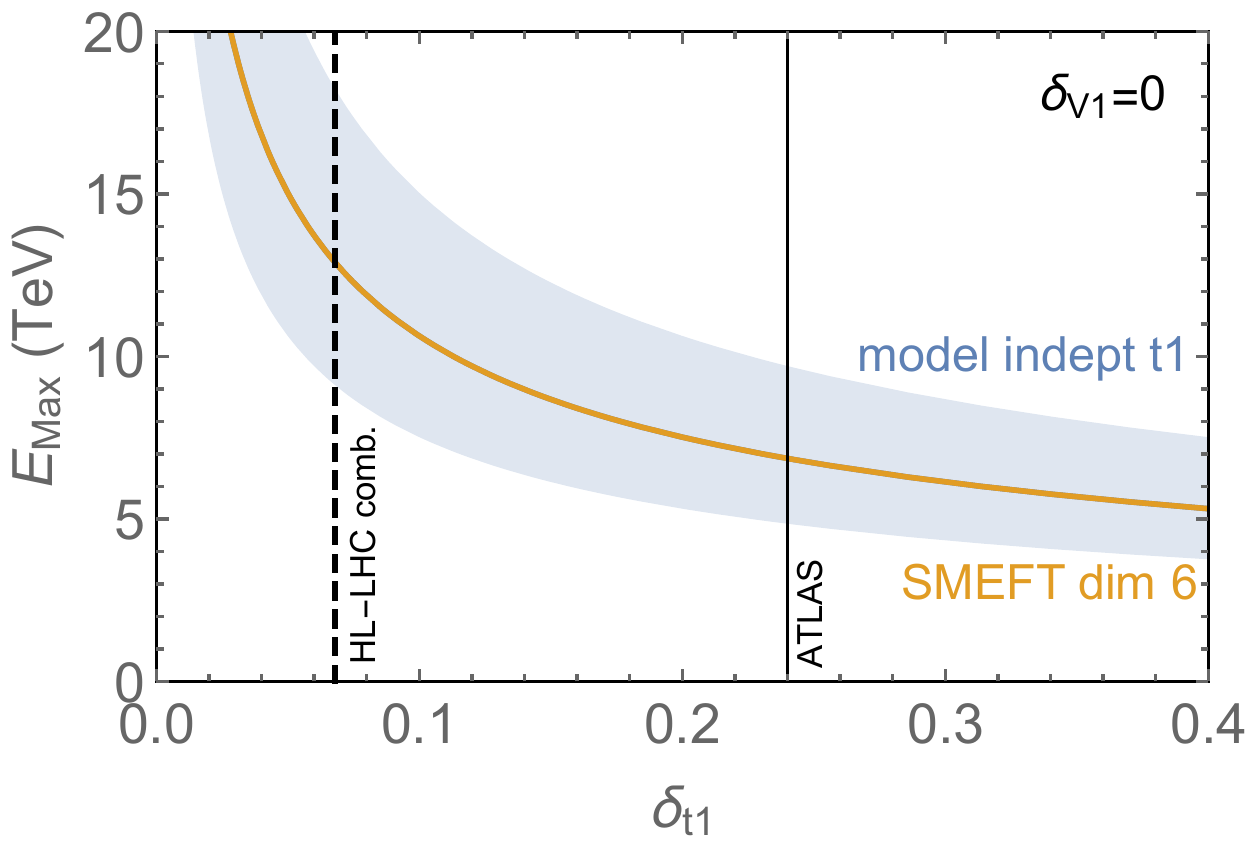}}
\caption{\small The unitarity bound on $\de_{t1}$ assuming
$\de_{W1}, \de_{Z1} = 0$.
The model-independent bound is equal to the optimal bound for all values
of $\de_{t1}$ shown.
The band around the model-independent scale results from varying the unitarity
bound to $\frac 12 \le |\hat{\scr{M}}| \le 2$.
For comparison, we show the 95\% C.L.~limits on the coupling from ATLAS \cite{atlas:2020atl} and a projected HL-LHC combination  \cite{Cepeda:2019klc}. \label{fig:SMEFTt1}}
\end{minipage}}
\end{figure}

\subsection{Optimal Bound}
To further discuss the implications of $\de_{t1}$,
we consider a scenario where $\de_{t1}$ is nonzero, but all the other Higgs
couplings are compatible with the SM.
To estimate the scale of new physics in this scenario, it is conservative to assume
$\de_{W1}, \de_{Z1} = 0$, since unitarity bounds from \Eq{VVVVboundsmodelindep} are stronger 
than \Eq{t1modelindpt}. 
As in previous sections, we consider the optimal bound obtained by marginalizing
over the infinitely many unmeasured couplings.
The optimal bound can be constrained by considering the SMEFT operator
\[
\eql{tthSMEFT}
\de\scr{L}_\text{SMEFT}
= \frac{y_t}{M^2} \left(H^\dagger H - \frac{v^2}{2}\right)
(\bar{Q}_L \tilde{H} t^{\vphantom\dagger}_R + \hc),
\]
which gives 
\[
\eql{SMEFTdeviation}
\de_{t1} = -\frac{v^2}{M^2}, \quad c_{t2} = c_{t3} = 3 \de_{t1},
\]
and $c_{tn} = 0$ for $n \ge 4$.
This imposes additional unitarity bounds.
We find that the bounds for the model-independent processes considered above
give the most stringent bound for small $\de_{t1}$, but for larger values
of $\de_{t1}$ the strongest bound comes from $\bar{t}_R t_R \to hh$, which gives
\[
E_\text{max} \simeq \frac{2.4\TeV}{|\de_{t1}|}.
\]
However, this only dominates over the bounds in \Eq{t1modelindpt} for
$\de_{t1} \gsim 0.6$, which is larger than allowed by current constraints.
In Fig.~\ref{fig:SMEFTt1} we show the unitarity bounds on $\de_{t1}$ along
with the experimental bounds from ATLAS and the projected sensitivity of a HL-LHC combination.

\begin{figure}[!t]
\centerline{\begin{minipage}{0.8\textwidth}
\centering
\centerline{\includegraphics[width=230pt]{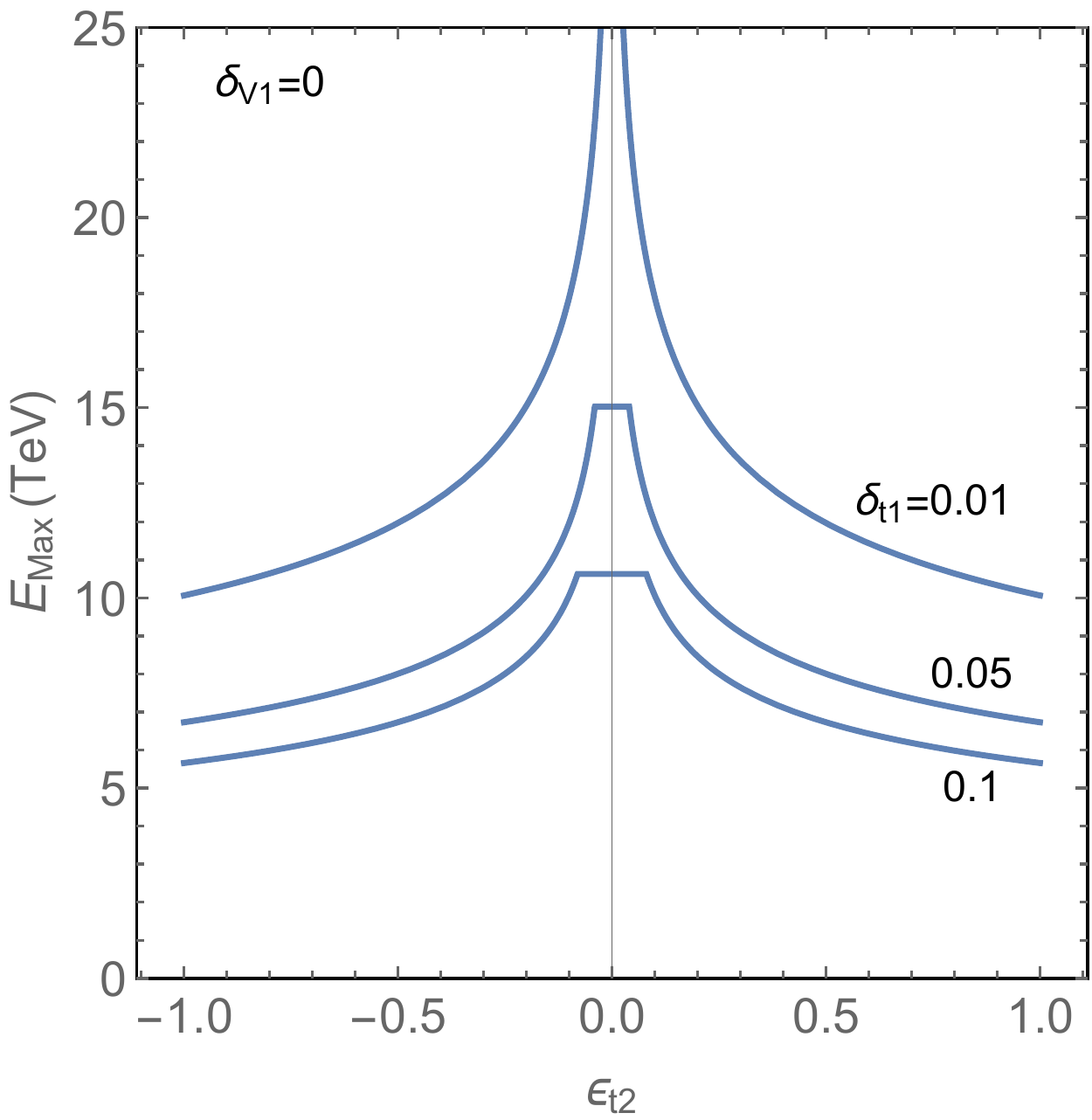}}
\caption{\small The unitarity bound from processes that depend on $\de_{t1}$, $c_{t2} = 3\de_{t1}(1+\epsilon_{t2})$ where $\epsilon_{t2}=0$ is the prediction of the dimension-6 SMEFT operator.  Due to these amplitudes depending on coupling $\delta_{V1}$, it has been set to zero in this plot.     
\label{fig:t1epsilon2}}
\end{minipage}}
\end{figure}

\subsection{SMEFT Predictions from Unitarity}
If the scale of new physics is high, we expect that an observed deviation
in the Higgs couplings can be described by the lowest-dimension SMEFT operator.
In the case of the $\bar{t}t h$ coupling, this is the operator given in \Eq{tthSMEFT},
which makes the predictions \Eq{SMEFTdeviation} for the higher-order deviations.
We can constrain the accuracy of these predictions from unitarity, as 
outlined in previous sections.
The results are shown in Fig.~\ref{fig:t1epsilon2}.
As expected, the SMEFT predictions are accurate only if the scale of
new physics is $\gsim 10\TeV$.

\section{New Physics from $hhVV$ and $hh\bar{t}t$ Couplings}
\label{sec:higgsquad}
In this section we discuss the implications of a deviation in the $hhVV$ or $hh\bar{t}t$ coupling,
parameterized respectively by $\de_{V2}$ and $c_{t2}$ in \Eq{theL}.
Since there are no symmetries to prevent this, any new physics that contributes to these couplings should  also
contribute to a comparable deviation in $\de_{V1}$ and $\de_{t1}$, which will be measured
to greater precision.
On the other hand, it is possible that $\de_{V1}$ and $\de_{t1}$ are suppressed by an
accidental cancellation.
In any case, experimental constraints on $\de_{V2}$ and $c_{t2}$ will improve
dramatically at the HL-LHC, and will give us additional information about possible
new physics.
Another motivation for studying these couplings is that they directly contribute
to di-Higgs production.
Therefore, an anomalous rate for di-Higgs production may be due to $\de_{V2}$ 
(in vector boson fusion) or $\de_{t2}$ (from gluon fusion).  Therefore we should consider these couplings in order to determine 
the unitarity bounds from any future di-Higgs anomalies.

\subsection{$hhVV$: Model-Independent Bound on the Scale of New Physics}
We now work out the model-independent bound on the scale of new physics coming
from an observation of $\de_{V2} \ne 0$.
This coupling can be measured from di-Higgs production via vector boson fusion
\cite{Bishara:2016kjn}.
Although this process in principle is sensitive to an anomaly in the $h^3$ 
coupling, this sensitivity is strongly reduced by requiring large di-Higgs
invariant mass to suppress backgrounds.
Because any new physics that contributes to $\de_{V2}$ will also contribute to
$\de_{V1}$, we assume that both couplings are nonzero in the present discussion.

\begin{figure}[!t]
\centerline{\begin{minipage}{0.8\textwidth}
\centering
\centerline{\includegraphics[width=300pt]{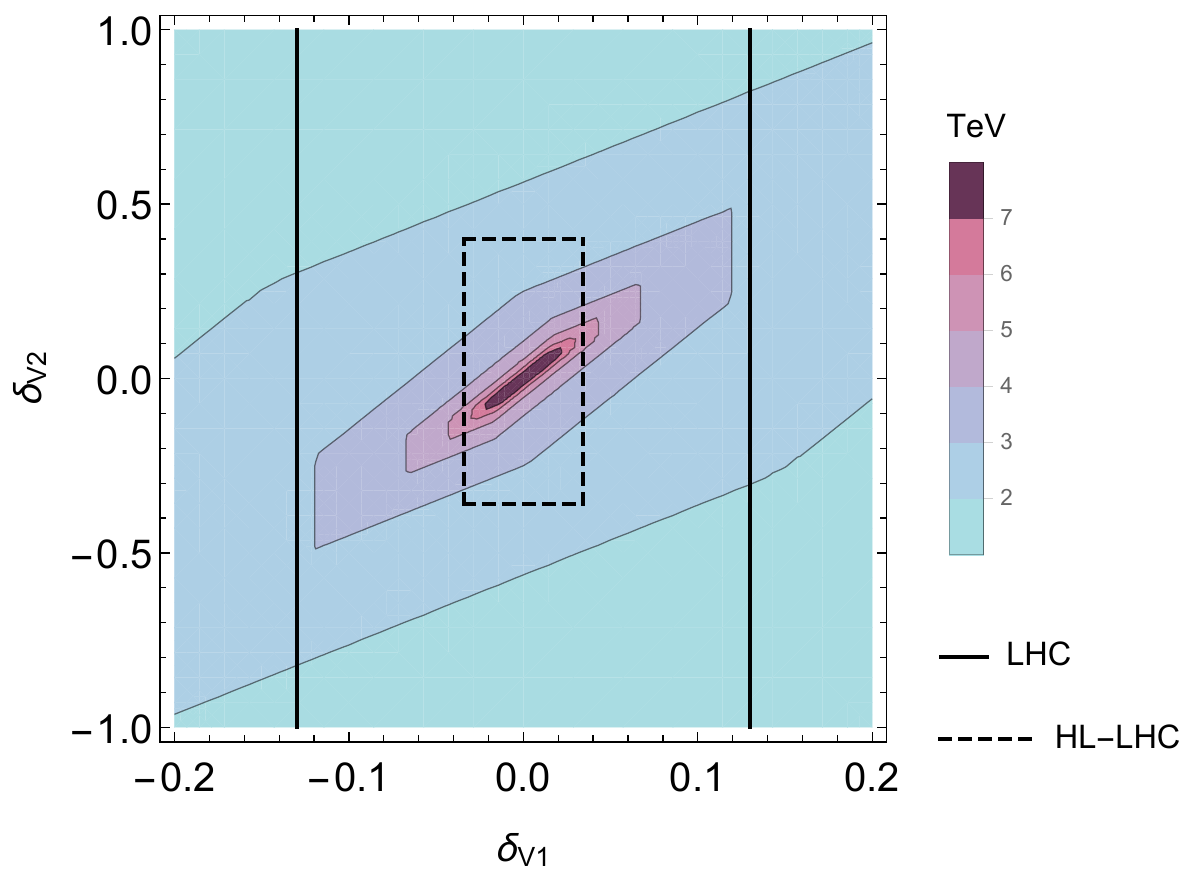}}
\caption{\small Unitarity violating contours from $\de_{V1}$ and $\de_{V2}$. The solid lines represent the ATLAS bound on $\de_{V1}$ \cite{atlas:2020atl} while the $\de_{V2}$ bound \cite{atlas:2020atl2} is outside of the plot range. The dashed lines show the projected bounds for $\de_{V1}$ \cite{Cepeda:2019klc} and $\de_{V2}$ at HL-LHC, where the $\de_{V2}$ bounds are the 95\% C.L.~bounds from doubling the 68\% bounds from a projected vector boson fusion di-Higgs search \cite{Bishara:2016kjn}. }
\label{fig:vbfnoiso}
\end{minipage}}
\end{figure}

The procedure we use to obtain the model-independent bound is an extension of the
one used in \S\ref{sec:higgshVV} to include $\de_{V2} \ne 0$.
This adds the model-independent processes  
$h^2 V_L^2$,
$h V_L^4$,
and $V_L^6$.
Because the $\de_{V1}$ and $\de_{V2}$ couplings each contain 2 derivatives
additional insertions of these vertices can cancel the $1/E^2$ from additional
propagators.
This means that the leading diagrams at high energy include diagrams with 
multiple propagators.
We find
\[
\eql{vvhhform}
\begin{split}
\hat{\scr{M}}(V_L V_L \to h h) 
&\sim \left(\de_{V1} + \de_{V2} + \de_{V1}^2\right)\frac{E^2}{v^2},\\
\hat{\scr{M}}(V_L V_L \to V_L V_L h)
&\sim \left(\de_{V1} + \de_{V2} + \de_{V1}^2 + \de_{V1}\de_{V2}
+ \de_{V1}^3\right)\frac{E^3}{v^3},
\\
\hat{\scr{M}}(V_L V_L V_L \to V_L V_L V_L)
&\sim\left(\de_{V1} + \de_{V2} + \de_{V1}^2 + \de_{V1} \de_{V2}
+ \de_{V1}^2 \de_{V2}
+ \de_{V1}^3 + \de_{V1}^4\right)\frac{E^4}{v^4}.
\end{split}
\]
Amplitudes related to these by crossing have the same scaling.
Current experimental constraints give $|\de_{V1}| \lsim 0.2$,
while $\de_{V2}$ has a weak constraint of $-1.8 \le \de_{V2} \le 1.9$ at $95\%$ C.L. \cite{atlas:2020atl2}.
We can therefore neglect the nonlinear terms in these
amplitudes (which are also much more difficult to compute).
Assuming custodial symmetry
($\de_{Z1} = \de_{W1}$, $\de_{Z2} = \de_{W2}$)
the strongest bounds are
\[
\begin{split}
 W^{+}_{L}W^{-}_{L} \to hh: \quad E_\text{max} \simeq \frac{1.5 \TeV}{{|\de_{V2}}-2\de_{V1}|^{1/2}}, \\
Z_L Z_L \to hW^{+}_{L}W_{L}^{-}: \quad E_\text{max} \simeq \frac{1.9 \TeV}{|\de_{V2}-4\de_{V1}|^{1/3}}, \\
W_{L}^{+}W_{L}^{+}Z_{L} \rightarrow W_{L}^{+}W_{L}^{+}Z_{L}: \quad E_\text{max} \simeq \frac{2.6\TeV}{|\de_{V2}-4\de_{V1}|^{1/4}}.
\eql{Vboundsv2}
\end{split}
\]
In Fig.~\ref{fig:vbfnoiso}, we show the unitarity violating scale given values of $\delta_{V1}$ and $\delta_{V2}$ along with the bounds on both coupling deviations from standard searches and a search for vector boson fusion di-Higgs.  The figure shows that HL-LHC searches for VBF di-Higgs could find coupling deviations with unitarity bounds below $3$ TeV.

\begin{figure}[!t]
\centerline{\begin{minipage}{0.8\textwidth}
\centering
\centerline{\includegraphics[width=250pt]{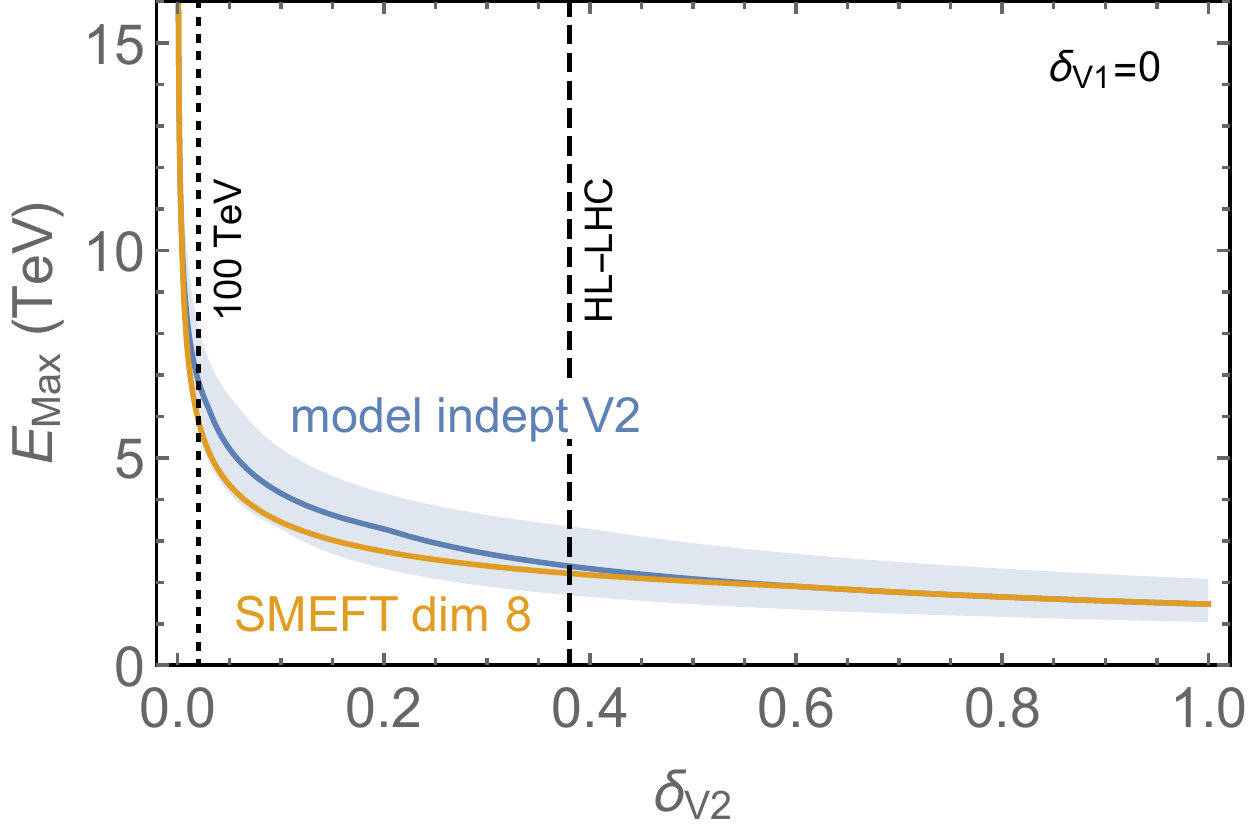}}
\caption{\small The unitarity bound from as a function of $\de_{V2}$
neglecting small terms proportional to $\de_{V1}$.
The optimal bound lies between the model-independent and SMEFT estimates.
The band around the model-independent bound results from varying the
unitarity bound to $\sfrac 12 \le |\hat{\scr{M}}| \le 2$.
For comparison, we show 95\% C.L.~limits on the coupling from the vector boson 
fusion di-Higgs analysis projected for the HL-LHC and a 100 TeV $pp$ collider \cite{Bishara:2016kjn}.}
\label{fig:SMEFTV2}
\end{minipage}}
\end{figure}

\subsection{$hhVV$: Optimal Bound and SMEFT Predictions}
We now consider the optimal bound obtained by marginalizing over the
infinitely many unmeasured couplings.
As in previous sections, we do this by considering a scenario where these
couplings are given by a single SMEFT operator.
In the present case, we use the dimension-8 operator
\[
\eql{SMEFThhVV}
\frac{1}{M^4}\left(H^\dagger H - \frac{v^2}{2}\right)^2 D^\mu H^\dagger D_\mu H,
\]
which gives custodial symmetry preserving couplings.
Performing field redefinitions to remove the Higgs self couplings at order $1/M^4$, 
we have find that the Higgs couplings to the vector bosons are given by 
\[
\eql{hhVVSMEFTrealization}
\de_{V1}=0,\quad \de_{V2}=\frac{v^4}{M^4}, \quad c_{V3} = 8 \de_{V2}, 
\quad 
c_{V4} = 32 \de_{V2}, 
\quad
c_{V5} = 72 \de_{V2},
\quad
c_{V6} = 72 \de_{V2},
\]
and $c_{Vn} = 0$ for $n \geq 7$.
The unitarity bound obtained from this operator is always stronger than the
optimal bound, so the optimal bound lies between this bound and the model-independent
bound computed above.
In Fig. \ref{fig:SMEFTV2}, we plot both the model-independent and the 
SMEFT unitarity bound as a function of $\de_{V2}$, neglecting terms proportional to 
$\de_{V1}$, showing that the optimal bound is close to the model-independent one.

\begin{figure}[!t]
\centerline{\begin{minipage}{0.8\textwidth}
\centering
\centerline{\includegraphics[width=250pt]{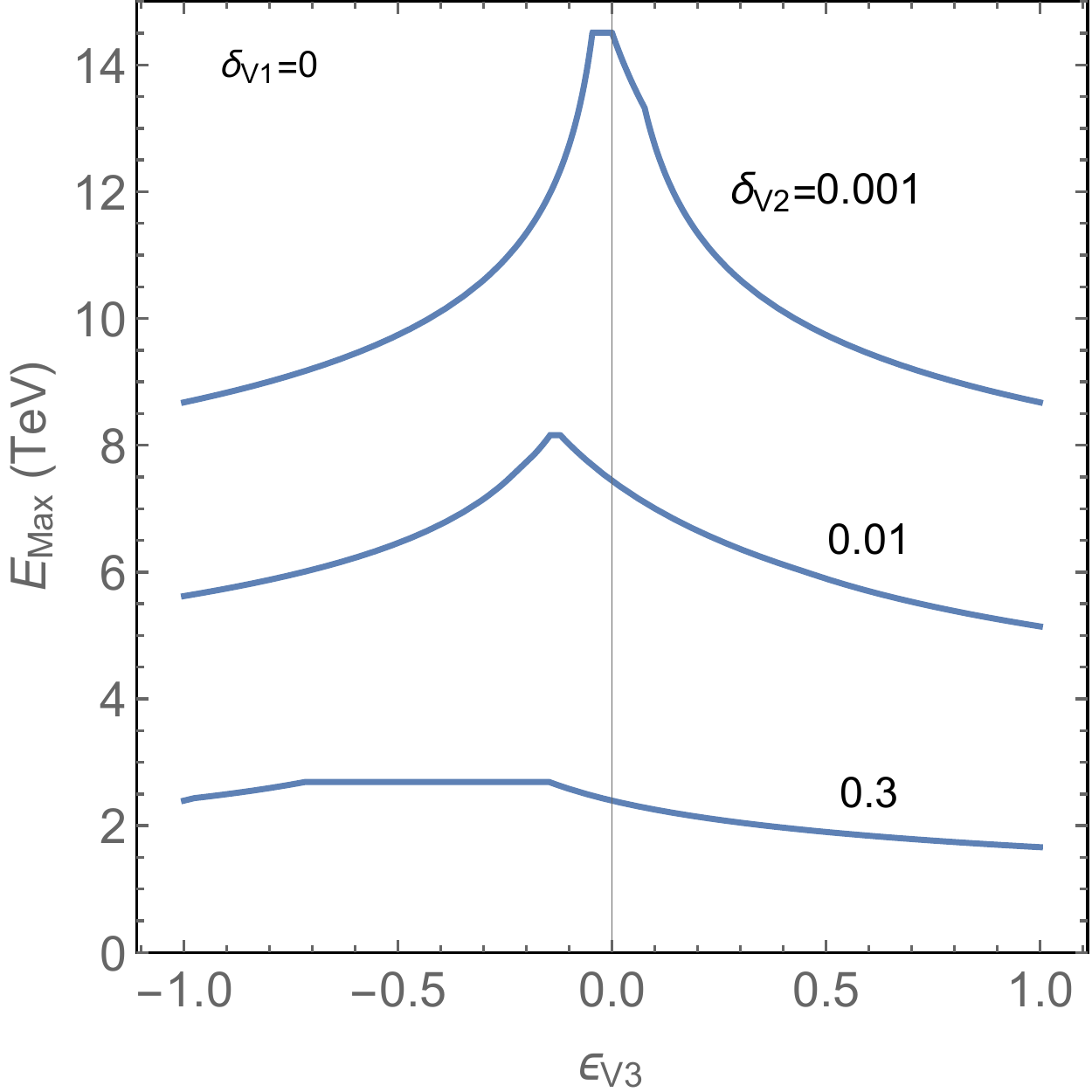}}
\caption{\small The unitarity bound from processes that depend on $\de_{V2}$ and $c_{V3} = 8\de_{V2}(1+\epsilon_{V3})$ to linear order, where $\epsilon_{V3} = 0$ correspond to the SMEFT predictions, assuming $\de_{V1}=0$.}
\label{fig:v2epsilon3}
\end{minipage}}
\end{figure}

Next, we consider the accuracy of the SMEFT prediction for $\de_{V2}$
from the operator \Eq{SMEFThhVV}.
(We again consider the case where $\de_{V1} = 0$).
We expect the predictions of this operator to become more accurate as the
scale of new physics becomes large.
In Fig.~\ref{fig:v2epsilon3} we plot the quantity
\[
\ep_{V3} = \frac{c_{V3} - c_{V3}^{\text{dim 8}}}{c_{V3}^{\text{dim 8}}},
\]
where $c_{V3}^{\text{dim 8}} = 8 \de_{V2}$.
As in previous cases, 
we find that the SMEFT prediction becomes accurate when the scale of new
physics is larger than a TeV.

\subsection{$hh\bar{t}t$: Model-Independent Bound on the Scale of New Physics}
We now consider a deviation in the $hh\bar{t}t$ coupling $c_{t2}$.
The study of this coupling is strongly motivated by the fact that di-Higgs
production is sensitive to this coupling, and therefore di-Higgs production
does not measure the $h^3$ coupling in a model-independent way \cite{Azatov:2015oxa}.
However,  measuring $ht\bar{t}$ and $hht\bar{t}$ production has been shown to break the degeneracies between the $hhh$, $h\bar{t}t$ and $hh\bar{t}t$ couplings  \cite{Englert:2014uqa, Li:2019uyy, taoliu:2019ijk}.

In this subsection we focus on the unitarity bound on $c_{t2}$.
We are interested in model-independent processes that do not depend on
$c_{tn}$ for $n \ge 3$.
The relevant couplings are given in \Eqs{ttstructure} and \eq{VVXMaster}.
We can work out that the model-independent processes have the schematic form at leading order in the energy expansion:
\[
\eql{tthhform}
\begin{split}
\hat{\scr{M}}(\bar{t} t \to hh) 
&\sim y_t c_{t2} \, \frac{E}{v},
\\
\hat{\scr{M}}(\bar{t} t \to V_L h h) 
&\sim y_t \bigl(\de_{t1}+c_{t2} + \de_{V1} + \de_{V2}+\de_{t1}\de_{V1}+\de_{V1}^2 \bigr)
\, \frac{E^2}{v^2},
\\
\hat{\scr{M}}(\bar{t} t \to V_L V_L h) 
&\sim y_t \bigl(\de_{t1} + c_{t2} + \de_{V1} + \de_{V2}
+ \de_{t1} \de_{V1}+\de_{t1} \de_{V2} 
\\[-2pt]
&\qquad\ \,{}
+ c_{t2} \de_{V1} + \de_{V1}^2 + \de_{t1} \de_{V1}^2\bigr)
\, \frac{E^2}{v^2},
\\
\hat{\scr{M}}(\bar{t} t \to V_L V_L V_L  h) 
&\sim y_t \bigl(\de_{t1} + c_{t2} + \de_{V1} + \de_{V2}
+ \de_{t1} \de_{V1}+\de_{t1} \de_{V2} 
\\[-2pt]
&\qquad\ \,{}
+ c_{t2} \de_{V1} + \de_{V1}^2 + \de_{V1} \de_{V2}
+ \de_{t1}\de_{V1}^2+\de_{V1}^3\bigr)
\, \frac{E^3}{v^3},
\\
\hat{\scr{M}}(\bar{t} t \to V_L V_L V_L V_L V_L) 
&\sim y_t \bigl(\de_{t1} + c_{t2} + \de_{V1} + \de_{V2}
+ \de_{t1} \de_{V1}+\de_{t1} \de_{V2} + c_{t2} \de_{V1} 
\\[3pt]
&\qquad\ \,{}
+ \de_{V1}^2 + \de_{V1} \de_{V2} +\de_{t1}\de_{V1}^2+\de_{t1}\de_{V1}\de_{V2}+c_{t2}\de_{V1}^2
\\[-2pt]
&\qquad\ \,{}
+ \de_{V1}^3+\de_{V1}^2\de_{V2}+\de_{t1}\de_{V1}^3+\de_{V1}^4)
\, \frac{E^4}{v^4}.
\end{split}
\]
For $\bar{t}b$ initial states, the first and third process vanish while the second process does not have a $\de_{t1}$ term. 
Amplitudes related to these by crossing have the same scaling.  Again, due to constraints on $\de_{t1}, \de_{V1}$ we can neglect the nonlinear  terms.  
At linear order, we see that only the $\bar{t} t \to hh$ amplitude is independent of $\de_{V2}$, 
which is poorly constrained experimentally and thus can substantially affect the constraints on $c_{t2}$.
These linear contributions involving $\de_{V1}$ and $\de_{V2}$ involve diagrams with
propagators, which are significantly more difficult to compute so we have focused on the terms from $\de_{V2}$.
%
%
Due to this contamination from $\de_{V2}$, we will use only $\bar{t} t \to hh$ to set unitarity bounds
on $c_{t2}$.
The bounds taking into account the dominant linear contributions are: 
\[
\begin{split}
t_{R}\bar{t}_{R}\rightarrow hh:& \quad E_\text{max} \simeq \frac{7.2 \TeV}{|c_{t2}|},\\
t_{R}\bar{t}_{R}\rightarrow W_{L}^{+}W_{L}^{-}h: & \quad E_\text{max} \simeq \frac{4.7 \TeV}{|c_{t2}-2\de_{t1}+\frac{1}{3}\de_{V2}|^{1/2}},  \\
\ t_{R}\bar{b}_{R}\rightarrow W_{L}^{+}h^2:& \quad E_\text{max} \simeq \frac{4.7 \TeV}{|c_{t2}-2\de_{t1}-\frac{2}{3}\de_{V2}|^{1/2}}, \\
t_{R}\bar{b}_{R}W_{L}^{-} \rightarrow hW_{L}^{+}W_{L}^{-}:& \quad E_\text{max} \simeq \frac{3.9 \TeV}{|c_{t2}-3\de_{t1}+\frac{1}{2}\de_{V2}|^{1/3}}, \\
t_{R}\bar{b}_{R}W_{L}^{-} \rightarrow W_{L}^{+}W_{L}^{+}W_{L}^{-}W_{L}^{-}:& \quad E_\text{max} \simeq \frac{4.2 \TeV}{|c_{t2}-3\de_{t1}+\frac{1}{3}\de_{V2}|^{1/4}}.\\
\eql{tprocessd2}
\end{split}
\]
\begin{figure}[!t]
\centerline{\begin{minipage}{0.8\textwidth}
\centering
\centerline{\includegraphics[width=300pt]{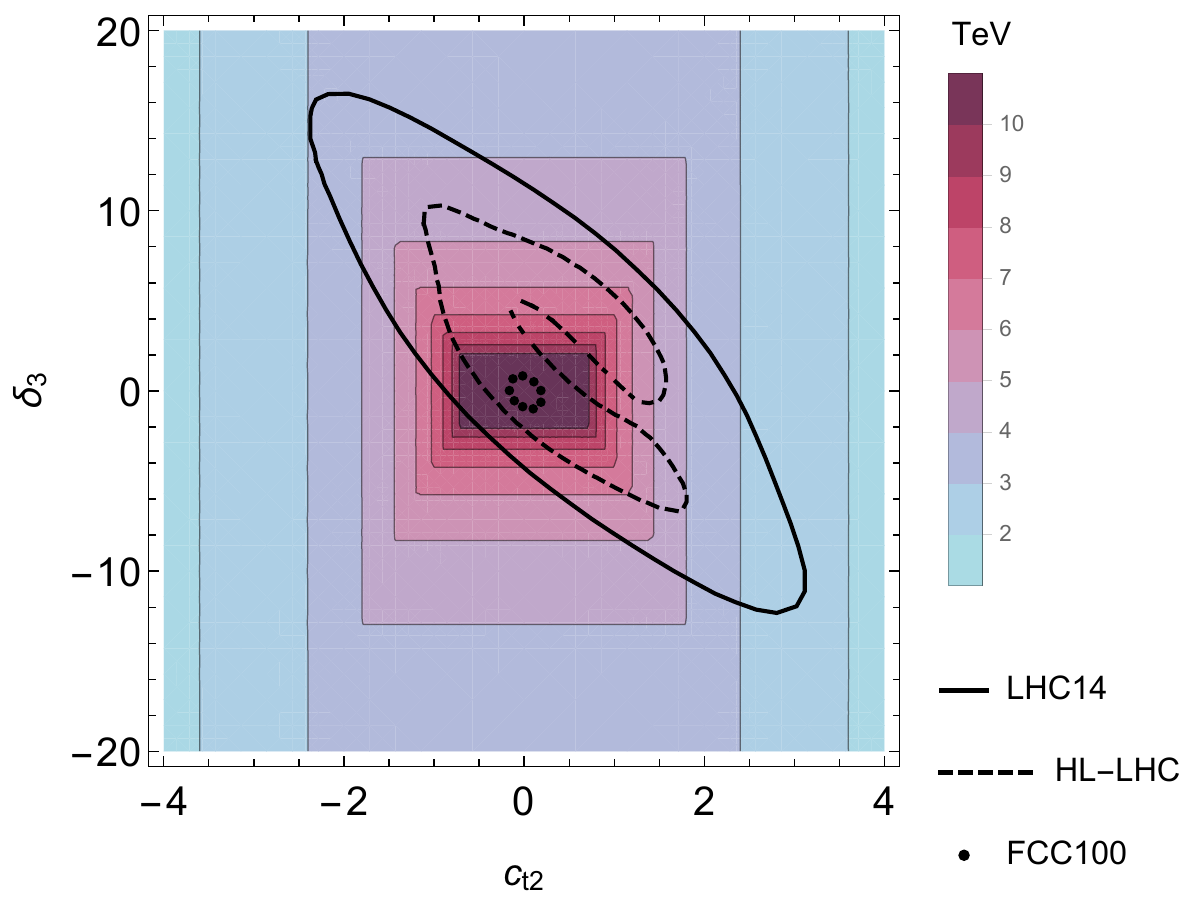}}
\caption{\small Unitarity violating contours from $\de_{3}$ and $c_{t2}$. The $95\%$ C.L.~ projections from gluon fusion di-Higgs searches are shown for the LHC (solid) and for the HL-LHC (dashed), which were obtained by expanding the $1\si$ contours of \cite{Azatov:2015oxa} by 1.6 to estimate the $95\%$ 
C.L.~sensitivity.}
\label{fig:ggfnoiso}
\end{minipage}}
\end{figure}

In Fig.~\ref{fig:ggfnoiso}, we plot the unitarity violating scale as a function
of $c_{t2}$ and $\de_3$.
Superimposed on the plot are estimates of the current bounds and sensitivity
to these parameters from gluon fusion di-Higgs production \cite{Azatov:2015oxa}.
We see that it is plausible that the HL-LHC could find deviations that point to a
scale of new physics below $3$ TeV, even allowing for the experimental
degeneracy between $c_{t2}$ and $\de_3$.

\subsection{$hh\bar{t}t$: Optimal Bound and SMEFT Predictions}
\begin{figure}[!t]
\centerline{\begin{minipage}{0.8\textwidth}
\centering
\centerline{\includegraphics[width=230pt]{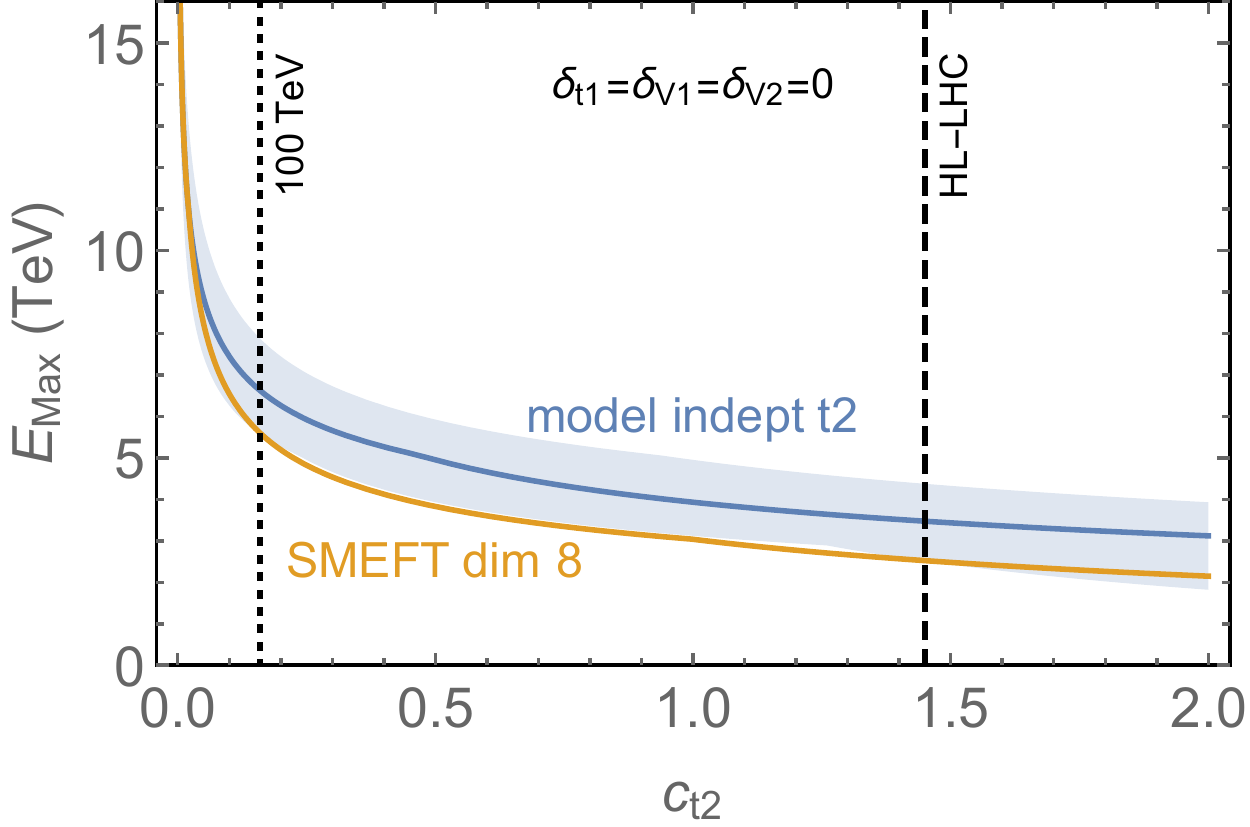}}
\caption{\small The unitarity bounds from both the model-independent approach and the SMEFT dimension-8 prediction, the optimized bound from marginalizing over other couplings should be somewhere between these two lines. We assume $\delta_{t1}=\delta_{V1}=\delta_{V2}=0$. We also plot the projected 95\% C.L.~limits on the coupling from the gluon fusion di-Higgs analysis at the  HL-LHC and a 100 TeV $pp$ collider \cite{Azatov:2015oxa}}
\label{fig:SMEFTt2}
\end{minipage}}
\end{figure}

To obtain the relations between $c_{2t}$ and higher order couplings, we use the dimension-8 SMEFT operator
\[
\frac{y_t}{M^4}
\left( H^\dagger H - \frac{v^2}{2} \right)^2(\bar{Q}_L \tilde{H} t^{\vphantom\dagger}_R + \hc),
\]
which gives the predictions
\[
\de_{t1}=0,\hspace{0.5 cm} c_{t2}=-2\frac{v^4}{M^4},\hspace{0.5 cm} c_{t3} = 6 c_{t2}, \hspace{0.5 cm} c_{t4} = 15 c_{t2}, \hspace{0.5 cm} c_{t5} = 15 c_{t2}, 
\eql{ct2SMEFTrealization}
\]
and $c_{tn} = 0$ for $n \geq 6$.
As in the previous cases, we can use \Eq{ct2SMEFTrealization} to obtain unitarity bounds from processes that we classified as model-independent. Fig \ref{fig:SMEFTt2} shows the unitarity bounds predicted by the model independent approach and the SMEFT operator, where we assume $\delta_{t1}=\delta_{V1}=\delta_{V2}=0$ to focus on $c_{t2}$.  Thus, the optimal bound is still within our estimated uncertainty of the model-independent bound.  

\begin{figure}[!t]
\centerline{\begin{minipage}{0.8\textwidth}
\centering
\centerline{\includegraphics[width=230pt]{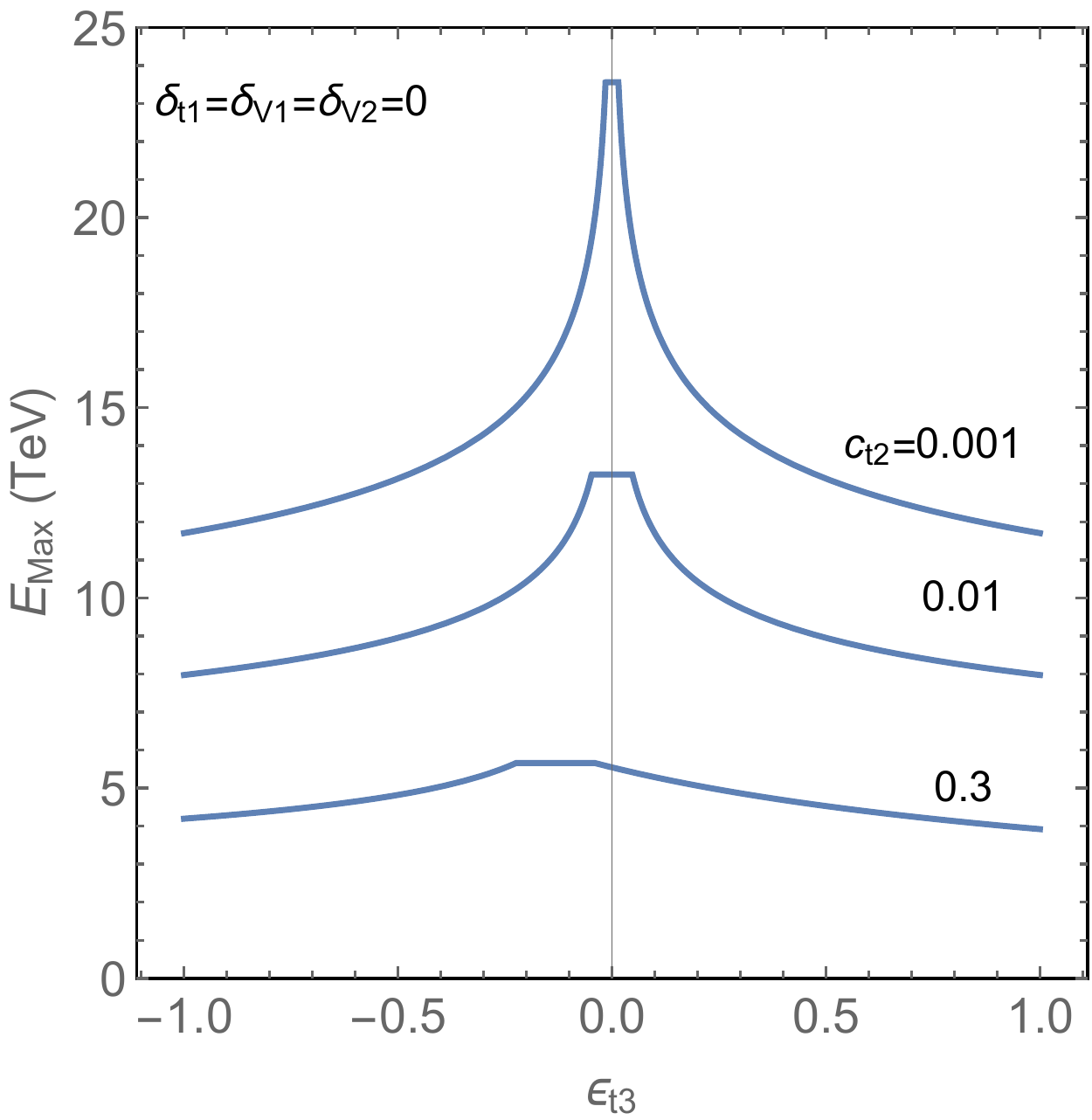}}
\caption{\small The unitarity bound from processes that depend only on $c_{t2}$ and $c_{t3} =6c_{t2}(1+\epsilon_{3})$ when $\de_{t1}=\de_{V1}=\de_{V2}=0$. Setting $\epsilon_{t3} = 0$ corresponds to the SMEFT prediction from the dimension-8 operator.}
\label{fig:t2epsilon3}
\end{minipage}}
\end{figure}

Once again, we can see the effect that a high scale of unitarity violation (compared to 1 TeV) has on the SMEFT predictions in \Eq{ct2SMEFTrealization}. Fig. \ref{fig:t2epsilon3} shows the unitarity scale dependence on $\epsilon_{t3}$ where $c_{t3}=6c_{t2}(1+\epsilon_{t3})$ and we assume $\delta_{t1}=\delta_{V1}=\delta_{V2}=0$. As with the other couplings, at high scales of unitarity violation (e.g. 10 TeV), $c_{t3}$ is close to its SMEFT value.

\section{Conclusions}
\label{sec:conclusions}
In this paper, we have investigated the scale of unitarity violation due to nonstandard Higgs self-couplings, and Higgs couplings to $W/Z$ bosons and top quarks.  
In the SM, good high energy behavior for multiparticle scattering amplitudes
relies on delicate cancellations among the various Higgs couplings. 
If these cancellations are upset by new physics contributions to the Higgs couplings, 
this leads to tree-level unitarity violation at high energies,
signaling the breakdown of perturbation theory and the onset of new physics.
In this way, we can give a model-independent bound on the scale of new physics
directly from any observed deviation from the SM prediction for Higgs couplings.

In this work, we focused on the couplings $h^3$, $h^4$, $hVV$, $h^2VV$, 
$h\bar{t}t$, and $h^2\bar{t}t$  where $V = W$ or $Z$, 
which will be probed at the HL-LHC and future colliders.
In the SM, these couplings are predicted at the percent level while current 
constraints are only at the $10\%$--$100\%$ level.
Upcoming experiments will significantly improve these constraints, giving 
many opportunities to discover physics beyond the SM.
Our work translates these searches into a direct probe of the scale of new physics.

For the $hVV, h\bar{t}t$ couplings, the current constraints 
allow coupling values that require new physics below $3$~TeV for $W/Z$ couplings,
and below $8$~TeV for the top coupling.  
The Higgs trilinear coupling is much more weakly constrained, 
allowing a scale of new physics as low as $4$~TeV.  
The couplings $hh\bar{t}t$ and $hhVV$ are of particular interest for  di-Higgs 
searches in gluon-fusion and vector boson fusion, and their constraints allow 
a scale of new physics as low as $2$~TeV.   
These results show that measurements of Higgs couplings can point to a scale
of new physics within the kinematic reach for HL-LHC and future colliders.

Unitarity bounds can also place indirect constraints on couplings that are
difficult to measure directly, such as the $h^4$ coupling.
For example if there is a nonstandard Higgs trilinear coupling, we show that to 
keep the new physics bound above 10 TeV, the quartic coupling must closely 
approximate the coupling correlation from the dimension-6 SMEFT operator 
$(H^\dagger H)^3$.
We present similar results for the $W/Z$ and top couplings as well.  
We emphasize that these predictions do not make any assumptions about
the smallness of higher-dimension operators, and rely only on unitarity.

Our main conclusion is that, from a purely data-driven viewpoint, our
current knowledge of the Higgs couplings allows new physics at the few TeV scale.
This scale will be extensively probed at the HL-LHC and future colliders, both through direct
searches and Higgs coupling measurements, and there is a great deal of room for
discovery in both types of analyses.  In particular, the scales probed by the upcoming HL-LHC are not sufficiently large that
we can confidently neglect higher-dimension operators in SMEFT.   We have therefore adopted a completely bottom-up and model-independent
approach to translating these measurements into direct statements about the scale 
of new physics.  
We hope that these results will be useful in interpreting and further motivating the precision study of the Higgs boson's properties.

\section*{Acknowledgments}
We thank
T.~Cohen, X.~Lu, and D.~Soper
for discussions.
The work of SC was supported in part by the U.S.~Department of Energy under Grant Number 
DE-SC0011640.
The work of ML and MC was supported in part by the U.S.~Department of Energy under grant 
DE-SC-0009999.  
The work of FA was supported by the OCEVU Labex 
(ANR-11-LABX-0060) and the A*MIDEX project (ANR-11-IDEX-0001-02) funded by the 
``Investissements d'Avenir'' French government program, managed by the ANR.

\startappendices
\setcounter{section}{0}
\section{Calculation Techniques and Results}
\label{sec:appendix}
In this appendix we define the multi-particle amplitudes we use to obtain
the unitarity bounds, explain how they are computed, discuss potential infrared enhancements, and give the results
of the calculations used in the main text.
We extend the results of \Ref{Chang:2019vez} to include fermions,
momentum-dependent couplings, and tree-level diagrams with propagators.
\subsection{Scalar Amplitudes}
We first discuss amplitudes involving only scalar fields, which includes
amplitudes with longitudinal $W$ and $Z$ bosons when we use the equivalence
theorem.
Given $r$ species of scalars $\phi_1, \ldots, \phi_r$ we define the states
\[
\eql{ScalarStates1}
\ket{P; k_1, \ldots, k_r} 
&\equiv C_{k_1, \ldots, k_r} 
\myint d^4 x\ggap e^{-i P \cdot x} \ggap
\phi^{(-)}_1(x)^{k_1} \cdots \phi^{(-)}_r(x)^{k_r} \ket{0}
\nn
&= C_{k_1, \ldots, k_r}
\myint d\Phi_k(P; p_1, \ldots, p_k) \ggap
\ket{\phi_1(p_1) \cdots \phi_r(p_k)}.
\]
Here $k_1, \ldots, k_r$ are non-negative integers that give the
number of each species of particle in the state,
$\phi_i^{(-)}$ is the negative frequency (creation operator)
part of the interaction picture field $\phi_i$,
$\ket{\phi_1(p_1) \cdots \phi_r(p_k)}$ 
is an ordinary $k$-particle state with $k = k_1 + \cdots + k_r$, and
\[
d\Phi_k(P; p_1, \ldots, p_k)
= \frac{d^3 p_1}{(2\pi)^3} \frac{1}{2E_1} 
\cdots \frac{d^3 p_k}{(2\pi)^3} \frac{1}{2E_k} 
\ggap (2\pi)^4 \de^4(p_1 + \cdots + p_k - P)
\]
is the Lorentz invariant $k$-body phase space.
These states are $s$-wave states defined by integrating $k$-particle
states over the full phase space.
The normalization of the states is chosen to be
\[
\eql{scalarstatenorm}
\braket{P'; k'}{P; k}
= (2\pi)^4 \de^4(P' - P) \ggap \de_{k'k},
\]
where we use the abbreviations
\[
\ket{P; k} &= \ket{P; k_1, \ldots, k_r},
\qquad
\de_{k'k} = \de_{k'_1 k_1} \cdots \de_{k'_r k_r},
\qquad
C_k = C_{k_1, \ldots, k_r}.
\]
The normalization constant is given by
\[
\eql{ScalarNormalizationConstant2}
\frac{1}{|C_{k}|^{2}}
= k_1 ! \cdots k_r ! \ggap \Phi_k(P),
\]
where
\[
\eql{phasespacevol}
\Phi_k(P) &= \myint d\Phi_k(P) 
= \frac{1}{8\pi \gap (k-1)! \gap (k-2)!} 
\left( \frac{E}{4\pi} \right)^{\! 2k-4},
\]
is the total volume of phase space for massless particles
with center of mass energy $E = \sqrt{P^2}$.

We then consider $S$-matrix elements between these states:
\[
\eql{ScalarAmplitude1}
\bra{P'; k'} T \ket{P; k}
= (2\pi)^4 \de^4(P'-P)
\hat{\scr{M}}(P; k_1, \ldots,  k_r \to k'_1, \ldots, k'_r),
\]
where $S = 1 + iT$.
The amplitude $\hat{\scr{M}}$ is Lorentz invariant and depends only on $P_\mu$,
so it is a function of $E$ only. With the normalization \Eq{scalarstatenorm},
unitarity of the $S$ matrix implies that these amplitudes satisfy%
\[
|\hat{\scr{M}}| \le 1.
\]
For non-forward amplitudes this follows directly from the unitarity
of the $S$-matrix.
For forward amplitudes ($k'_i = k_i$) a few additional steps are required
to show that this holds for tree-level amplitudes, see \Ref{Chang:2019vez}.
This is the unitarity constraint we employ in this paper.

The Feynman rules for these amplitudes follow straightforwardly
from the standard rules.
The result is that the amplitude $\hat{\scr{M}}$ are obtained from the
standard Lorentz invariant amplitude $\scr{M}$ by averaging over the
initial and final state phase space:
\[
\eql{Mhatmaster}
\hat{\scr{M}}_{fi}(P) = C_f^* C\sub{i} \myint d\Phi_f(P) \ggap d\Phi_i(P) \ggap
\scr{M}_{fi},
\]
where $\scr{M}_{fi}$ is the usual Lorentz-invariant amplitude.%
\footnote{In more detail, \Eq{Mhatmaster} is
\[
\hat{\scr M}(P; k_1, \ldots, k_r \to k'_1, \ldots, k'_r)
&= C^*_{k'} C\sub{k} 
\myint d\Phi_{k'}(P; p'_1, \ldots, p'_{k'})
\ggap d\Phi_k(P; p_1, \ldots, p_k)
\nn
&\qquad\qquad\qquad\quad{} \times
\ggap \scr{M}(\phi_1(p_1) \cdots \phi_r(p_k) \to 
\phi_1(p'_1) \cdots \phi_r(p'_{k'})).
\]
}
Because we are averaging over final state momenta, these amplitudes have
contributions from disconnected diagrams, with each disconnected component contributing a $\hat{\scr{M}}$ factor, leading to a form $\hat{\scr{M}} \propto \Pi_{i} \hat{\scr{M}}_i$ .
However, the leading contribution to high-energy
amplitudes always comes from connected diagrams.

In simple cases, these amplitudes can be computed in terms of the
total volume of phase space given in \Eq{phasespacevol}.
For example, for a single insertion of a coupling with no derivatives
we have
\[
\frac{\displaystyle \bra{P'; k'} \myint d^4 x \ggap 
\phi_1^{n_1}(x) \cdots \phi_r^{n_r}(x) \ket{P; k}}{(2\pi)^4 \de^4(P' - P)}
&= C_{k'}^* C_k 
\ggap n_1! \cdots n_r! \Phi_{k'}(P) \Phi_{k}(P)\\ & = \frac{1}{C_{k'} C_k^*} 
\ggap \frac{n_1! \cdots n_r!}{k_1! \cdots k_r! k'_1! \cdots k'_r!} ,
\]
where we assume $n_i = k_i + k'_i$. 
For diagrams with a single insertion of a vertex containing derivatives,
we use the identities
\[
\eql{avP}
\myint d\Phi_k(P; p_1, \ldots, p_k) \ggap p_1^\mu
&= \frac{P^\mu}{k} \Phi_k(P),
\\
\myint d\Phi_k(P; p_1, \ldots, p_k) \ggap p_1 \cdot p_2
&= \frac{P^2}{2{ k \choose 2}} \Phi_k(P),
\]
which hold for the case where all particles are massless.

\subsection{States with One Fermion}
We consider a state containing a single fermion and $k$ scalars
\[
\eql{fermion1}
\ket{P; k_1, \ldots, k_r, \al, a} &\equiv C'_{k} \myint d^4 x \ggap
e^{-i P \cdot x} \ggap
\phi^{(-)}_1(x)^{k_1} \cdots \phi^{(-)}_r(x)^{k_r} 
\psi^{a\gap (-)}_{L\gap\alpha}(x)\ket{0}
\nn
&= C'_k \myint d\Phi_{k+1}(P; p_1, \ldots, p_k, q) \ggap
v^\al_L(q)
\ket{\phi_1(p_1) \cdots \phi_r(p_k) \widebar{\psi}_R^a(q)},
\]
where $\psi_L$ is a left-handed Weyl spinor field,
$\alpha$ is a spinor index,
and $a$ is a gauge index ({\it e.g.}~a color index).
Note that these states are given by phase space integrals of scattering
states weighted by a spinor wavefunction, so \Eq{Mhatmaster} is modified for
amplitudes involving these states.
(In the example above, the state created by the left-handed spinor field
is a right-handed antifermion.)
The normalization of these states is given by
\[
\braket{P'; k, \be, b}{P; k, \alpha, a} &= (2\pi)^4 \de^4(P' - P) 
\ggap k_1! \cdots k_r! \gap |C'_{k}|^2 \myint d\Phi_{k+1}(P; p_1, \ldots, p_k, q) \ggap
q^\mu \si^{\al\dot\be}_\mu \delta_{ab}
\nn
&= (2\pi)^4 \de^4(P' - P) 
\ggap k_1! \cdots k_r! \gap |C'_{k}|^2 \delta_{ab} \frac{P \cdot \si^{\al\dot\be}}{k+1}
\Phi_{k+1},
\]
where we used \Eq{avP}.
We choose the states \Eq{fermion1} to have normalization
\[
\braket{P'; k', \be, b}{P; k, \al, a} &= (2\pi)^4 \de^4(P' - P) \ggap
\delta_{ab} \de_{k'k} \frac{P \cdot \si^{\al\dot\be}}{E}.
\]
Note that in the $P^\mu$ rest frame we have
$P \cdot \si^{\al\dot\be} / E = \de^{\al\dot\be}$,
so this is the natural generalization of the normalization condition
\Eq{scalarstatenorm}.
The normalization constants are therefore given by
\[
\frac{1}{|C'_{k}|^2} = k_1! \cdots k_r! \ggap \frac{E}{k+1} \Phi_{k+1}(P).
\]

\subsection{States with Two Fermions}
We now consider states with two fermions and $k$ scalars of the form
\[
\ket{P; k_1, \ldots, k_r, L/R} &\equiv C''_{k}  \int d^4 x\, 
e^{-i P\cdot x} \ggap
\phi^{(-)}_1(x)^{k_1} \cdots \phi^{(-)}_r(x)^{k_r}
\widebar{\psi}^{a(-)}_{\gap R/L}(x) \psi^{a(-)}_{L/R}(x) \ket{0}
\nn
&= C''_k \myint d\Phi_{k+2}(P; p_1, \ldots, p_k, q, q') \ggap
\widebar{u}_{R/L}(q') v_{L/R}(q) \ggap
\nn
&\qquad\qquad{} \times
\sum_a \ket{\phi_1(p_1) \cdots  \phi_r(p_k) \psi^a_{R/L}(q') \widebar{\psi}^a_{R/L}(q)},
\]
where $\psi_L$ ($\psi_R$) are left-handed (right-handed) Weyl spinors.
In the massless limit the states $|\ldots L\rangle$ and $|\ldots R\rangle$ are 
orthogonal $s$-wave states, with the $L$ ($R$) state containing a 
fermion-antifermion pair which are both right-handed (left-handed) in helicity.    
These states are normalized as in \Eq{scalarstatenorm}
if we choose 
\[
\frac{1}{|C''_{k}|^2} = \gap k_1! \cdots k_r! \ggap
\frac{2N E^2}{(k+1)(k+2)} \Phi_{k+2}(P),
\]
where $a = 1, \ldots, N$ and for a top quark, $N=N_c$.
To compute amplitudes for these states, we use
\[
&\!\!\!\!\!
\frac{\displaystyle
\bra{P'; k'} \myint d^4 x\ggap \phi_1(x)^{n_1} \cdots \phi_r(x)^{n_r} 
\widebar{\psi}_{L/R}(x) \psi_{R/L}(x) \ket{P; k, L/R}}
{(2\pi)^4 \de^4(P' - P)}
\nonumber\\[5pt]
&\qquad\qquad\qquad\qquad\qquad{}
=  C_{k'}^* C''_{k} \ggap n_1! \cdots n_r! \ggap \frac{2N E^2}{(k+2)(k+1)} \Phi_{k'}(P) \Phi_{k+2}(P),
\nonumber\\[5pt]
&\qquad\qquad\qquad\qquad\qquad{}
= \frac{1}{C_{k'} (C''_k)^*} 
\ggap \frac{n_1! \cdots n_r!}{k_1! \cdots k_r! k'_1! \cdots k'_r!}, \\
&\frac{\displaystyle
\bra{P'; k'} \myint d^4 x\ggap \phi_1(x)^{n_1} \cdots \phi_r(x)^{n_r} 
\widebar{\psi}_{L/R}(x) \psi_{R/L}(x) \ket{P; k, R/L}}
{(2\pi)^4 \de^4(P' - P)} 
= 0.
\]

\subsection{Example Calculations}
We now give some examples of calculations involving these rules.
The amplitudes involving a single insertion of a vertex without derivatives
is straightforward using the formulas given above, and will not be discussed
further.
Diagrams with derivatives are less trivial because the derivatives may act
on fields that are connected with either initial or final state particles.
For example, consider
\[
\frac{\displaystyle
\bra{P'; 2} \myint d^4 x \ggap \phi^2 (\d\phi)^2 \ket{P; 2}}
{(2\pi)^4 \de^4(P' - P)}
&= \myint d^4 x \Bigl[
\bra{P'; 2} \phi^2 \ket{0} \bra{0} (\d\phi)^2 \ket{P;2}
\nonumber\\[-2pt]
&\qquad\qquad\ {}
+ \bra{P'; 2} (\d\phi)^2 \ket{0} \bra{0} \phi^2 \ket{P;2}
\nn
&\qquad\qquad\ {}
+ 4 \bra{P'; 2} \phi \d^\mu \phi \ket{0} \bra{0} \phi \d_\mu \phi \ket{P;2}
\Bigr]
\nn
&= 4 |C_2|^2  \left(
-2 \cdot \frac{1}{2} \ggap E^2 + 4 \cdot \ggap \frac{-iP^\mu}{2} \frac{iP_\mu}{2} \right) \Phi_2(P)^2
= 0.
\]
The cancellation can be understood at the level of the ordinary amplitude
from the fact that crossing symmetry implies that the amplitude is proportional
to $s + t + u = 4m_\phi^2$, which vanishes in the massless limit.

We now give an example of a diagram that contains a propagator:
\[
&\frac{\displaystyle
\bra{P'; 0, 0, 2} \myint d^4 x\ggap (\d\phi_3)^2 \phi_2 
\myint d^4 y\ggap \phi_2 (\d\phi_1)^2 \ket{P'; 2, 0, 0}}
{(2\pi)^4 \de^4(P' - P)}
\nn
&\qquad\qquad\qquad{}
= |C_2|^2 \myint d\Phi_2(P'; p_1', p_2') \ggap d\Phi_2(P; p_1, p_2) \ggap
(2p_1' \cdot p_2') (2 p_1 \cdot p_2) \frac{i}{P^2}
\nn
&\qquad\qquad\qquad{}
= |C_2|^2 \frac{i}{E^2} \bigl[\ggap E^2 \Phi_2(P) \bigr]^2.
\]
Diagrams with propagators are generally subleading at high energies compared
to diagrams with a single insertion.
There are a few relevant exceptions, which are discussed in the main paper.

\subsection{IR Enhancement}
The amplitudes $\hat{\scr{M}}$ are dimensionless, and once coupling constants
have been factored out, they depend on a single dimensionful variable $E$ 
in the massless limit.
The dependence on $E$ is therefore determined by dimensional analysis,
provided that there are no IR enhancements in the massless limit.
Such IR enhancements can arise because the integration over initial and final
state phase space can go over regions where internal propagators go on shell.
We now present arguments that such IR enhancements do not invalidate the
leading large $E$ scaling for any of the processes used to set the unitarity
bounds in this paper.
First, we show that many (but not all) possible IR enhancements 
can be ruled out by a simple parametric argument.
Second, 
we  give a diagrammatic argument that IR
enhancements can modify the \naive\ power counting by at most
corrections of order $\log(E/m)^n$ for some positive integer $n$,
where $m$ is the mass of a SM particle such as $m_W$ or $m_h$.
Finally, we point out that the gauge boson equivalence theorem itself is invalid in
the phase space region of the potential IR enhancements, since these are regions
where some Lorentz invariants $p_i \cdot p_j \sim m_W^2$ rather than $E^2$.
Therefore, phase space integration over these regions is suspect.
(We note that this issue arises already for $2 \to 2$ partial wave amplitudes.)
We argue that, because the singular phase space regions are parametrically small,
they cannot give rise to additional $\log(E/m_W)$ enhancements,
and therefore the Goldstone amplitudes correctly give the correct leading
behavior at large $E$.

For the parametric argument, consider an amplitude with leading large-$E$ behavior
\[
\eql{IRenhance}
\hat{\scr{M}} \sim C \left( \frac{E}{v} \right)^{\! n} 
\left( \frac{E}{m} \right)^{\! r} \log(E/m)^s,
\]
where $C$ is a BSM coupling, $m$ is an IR mass (such as $m_W$ or $m_h$),
and $n$, $r$, $s$ are non-negative integers.
Observe that if $r + s > 0$ this becomes arbitrarily large for any fixed $E$ 
in the limit  $m \to 0$ with $v$ and $c$ fixed.
But the amplitude cannot become arbitrarily large in this limit because
the massless limit is equivalent to a weak-coupling limit where the SM couplings
$g, \la, y_t \to 0$.
The coupling $C$ is held fixed in this limit, but can be chosen to be
arbitrarily small.
It is clear that we cannot have unitarity violation at arbitrary
energy scales in this limit, so IR enhancements of the form \Eq{IRenhance} 
are ruled out.

Note that the combinations $\la \de_3$, $\la \de_4$, $\la c_n$,
$y_t \de_{t1}$, and $y_t c_{tn}$ 
should be viewed as BSM couplings that are held fixed in the limit 
$\la, y_t \to 0$.
On the other hand, the couplings $\de_{V1}$, $\de_{V2}$,
and $c_{Vn}$ for $n \ge 3$ should be held fixed in the $g \to 0$ limit,
since these give Nambu-Goldstone interactions of finite strength in this limit.
This limit rules out many possible IR enhancements, but it is not
sufficient to justify the power counting of the amplitudes 
in \Eqs{hhhform}, \eq{vvhform}, \eq{tthform}, \eq{vvhhform}, and \eq{tthhform}.
In particular, it does not rule out power IR enhancements proportional to
additional powers of the SM couplings $g, \la, y_t$, for example
\[
\la \frac{E^2}{m_h^2} \sim \frac{E^2}{v^2},
\qquad
y_t \frac{E}{m_t} \sim \frac{E}{v}, \qquad g^2 \frac{E^2}{m_W^2}\sim \frac{E^2}{v^2},
\]
which have a finite weak-coupling limit as well as log terms such as
\[
\la \ln(E^2/m_h^2),
\qquad
y_t \ln(E/m_t),
\qquad g^2 \ln(E^2/m_W^2),
\]
which go to zero as  $\lambda,  y_t, g \to 0$.

Next, by examining the structure of the exchange diagrams, we will now argue that the IR enhancement of tree diagrams is
at most logarithmic.
In all the  amplitudes we computed, we find that such logs are
absent, although they may well be present in more complicated diagrams
that we have not computed.
As we point out below, even though the equivalence theorem cannot be trusted in parts of the phase space where the IR enhancement occurs, it is valid for a parametrically large region that could contribute to a logarithmic enhancement.  Therefore, the absence of logs in our calculations prove that the corresponding longitudinal gauge boson scattering amplitudes are free of logs.
By excising the small untrustworthy regions, we will then argue that the Nambu-Goldstone amplitudes can be used to set a conservative limit on the unitarity violating scale.
A better theoretical understanding of these log corrections is desirable,
but we will leave this for future work.

We now consider possible IR enhancements from a general tree
diagram contributing to the integrated amplitude $\hat{\scr{M}}$,
whether computed in the full SM or using the equivalence theorem.
An IR divergence can arise only from integrating over a region where an 
internal propagator becomes large.
This can happen if the momentum flowing through an internal line goes
on shell, or is soft.
If only a single propagator goes on shell, it is easy to understand
why the correction is at most logarithmic.
Consider an internal line with momentum $q - q'$, where 
$q$ ($q'$) is the momentum
of one of the initial (final) state particles.
Then the relevant part of the phase space integral is (in the massless limit)
\[
d^4 q \ggap \de(q^2) \ggap d^4 q' \ggap \de(q'^2) \ggap
\frac{1}{(q - q')^2}
&\propto 
\frac{d|\pvec{q}| \ggap d|\pvec{q}'| d\cos\th}{1 - \cos\th},
\]
where $\th$ is the angle between $\vec{q}$ and $\pvec{q}'$.
This integral diverges at most logarithmically because the integral
has a simple pole in $\cos\th$, which is one of the integration variables.  A general propagator with more legs attached can be analyzed by considering the following  momenta structure $P_1 + P_2 \to K_1 + K_2$ where  $P_1=(p_1+\cdots +p_r), P_2=(p_{r+1}+\cdots + p_n), K_1= (k_1+\cdots + k_s), K_2=(k_{s+1}+\cdots + k_m)$ and the momentum flowing through the propagator is $K_1 - P_1$.  By factorizing the incoming $n-$body phase into $r+(n-r)$-body phase space and similarly for the outgoing, we also see this propagator gives a log when integrating over $\cos \theta = \vec{P_1}\cdot \vec{K_1}/(| \vec{P_1}||\vec{K_1}|)$.

Next, we have to consider regions of the phase space integration where more
than one propagator gets large at the same time.
In all the cases we studied, the denominator of each of the large propagators
has a linear zero that depends on an independent parameter, either another angle or invariant mass of a set of particles,  that is integrated
over.  That is, near the singularity the integral behaves like
$\int dx \gap dy / x y$ and not $\int dx / x^2$.    
We checked this for $2 \to 2$ and $2 \to 3$ topologies, but we do 
not have a general proof for all topologies.  However, this makes intuitive sense given that a set of $n$ internal propagators going onshell requires $n$ independent conditions on the phase space.  Integrating over each of these conditions, then gives at most a
$\log^n(E/m)$ singularity.%
\footnote{In \Ref{Cornwall:1973tb} it is stated without proof that the $2 \to n$ partial wave
amplitudes have at most logarithmic singularities.}

We now note that in cases where there is a log enhancement in an amplitude involving longitudinal
gauge bosons, it is not obvious  whether the corresponding Nambu-Goldstone amplitude correctly 
reproduces these logs.
The gauge boson equivalence theorem guarantees that  the Nambu-Goldstone
amplitude correctly reproduces the full amplitude if 
$|p_i \cdot p_j| \gg m_V^2$ for all external 4-momenta $p_i$ and more generally for all Mandelstam invariants.
To see this, compare the exact dot products of longitudinal polarization vectors
\[
\ep_L(p_1) \cdot \ep_L(p_2) 
= \frac{E_1 E_2}{m_V^2} \left( \frac{|\vec{p}_1| |\vec{p}_2|}{E_1 E_2}
- \cos\th \right)
\]
with the approximation $\ep^{\mu}_{L}(p) \simeq p^\mu/m_V$:
\[
\frac{p_1}{m_V} \cdot \frac{p_2}{m_V}
= \frac{E_1 E_2}{m_V^2} 
\left( 1 - \frac{|\vec{p}_1| |\vec{p}_2|}{E_1 E_2} \cos\th \right), 
\]
where $\th$ is the angle between $\vec{p}_1$ and $\vec{p}_2$.
For $E_{1,2} \gg m_V^2$ and $\cos \th \ll 1$, these are equal up to 
corrections suppressed by $m_V^2/E^2$.
But for $\th \sim m_V/E$, the dot products are completely
different.
(For $\th = 0$, they even have opposite sign.)
This means that we cannot expect the equivalence theorem to be correct
in regions where some of the Mandelstam invariants are small.

This is relevant for the present discussion because these regions are precisely
the ones where one or more internal propagators can go on shell in the
massless limit, potentially giving an IR enhancement.
However, we note that the regions where the gauge boson equivalence theorem
does not apply are a parametrically small part of the phase space integral.
Integrals over such regions cannot give rise to IR singularities
of the form $\log(E/m)$, which instead arise from integrals of the form
$\sim \int dx/x$ over a parametrically large range $\De x \sim E/m$.
Thus, for example, when we obtain a Goldstone amplitude $\hat{M}$ that does
not have a $\log(E/m)$ enhancement, we know that the corresponding
gauge boson amplitude also does not have such an IR enhancement.
Omitting the singular region from the phase space integral in a Goldstone amplitude
without a log IR enhancement only changes the answer by a small correction
suppressed by powers of $m_W/E$, and therefore gives a good approximation
to the exact amplitude.

The discussion above has been less systematic than we would like.
It would be nice to have a better understanding of the gauge
boson equivalence theorem for partial wave amplitudes, including the
IR enhancements and subleading contributions.
We leave this for future work.

\subsection{Results}
We now give the results for the leading high-energy behavior for the processes
used in the main text in tables \ref{tab:a}-\ref{tab:j}.
All gauge bosons are understood to be longitudinally polarized. 
Also, note that since $Z_{L}$ is CP-odd, amplitudes involving an odd number of $Z_{L}$'s will be purely imaginary, however, these amplitudes can be made real by redefining the $Z_L$ states. All other processes are related to the ones listed in the tables via charge conjugation and/or crossing symmetry.  All of these amplitudes are calculated in the contact approximation.  As \Eqs{vvhform}, \eq{tthform}, \eq{vvhhform}, and \eq{tthhform} show, the nonlinear terms are small due to constraints on $\de_{V1}, \de_{t1}$.  However, there are linear terms proportional to $\de_{V1}, \de_{V2}$ in the top processes \Eqs{tthform} and \eq{tthhform}, so we've calculated the largest terms as shown in \Eqs{t1modelindpt} and \eq{tprocessd2}.

\begin{table}[!ht]
\centerline{\begin{minipage}{0.8\textwidth}
\centering
\vspace{1 mm}
\tabcolsep7pt\begin{tabular}{|c|c||c|c|}
\hline
 \textbf{Process} & \textbf{$\times\frac{\de_{V1}E^2}{8\pi v^2}$} &  \textbf{Process} & \textbf{$\times\frac{(\de_{V1}-\frac{1}{2}\de_{V2})E^2}{8\pi v^2}$} \\
\hline
$ZZ \rightarrow W^{+}W^{-}$ & $-\sqrt{2}$ & $hZ \rightarrow hZ$ & $-1$\\ 
$W^{+}W^{+} \rightarrow W^{+}W^{+}$ & 1& $ZZ \rightarrow hh$& $1$\\
$ZW^{+} \rightarrow ZW^{+}$ & 1& $hW^{+} \rightarrow hW^{+}$& $-1$\\
$W^{+}W^{-} \rightarrow W^{+}W^{-}$ & $-1$& $hh \rightarrow W^{+}W^{-}$& $\sqrt{2}$\\
\hline
\end{tabular}
\vspace{1 mm}
\caption{\label{tab:a} \small 4-body model-independent unitarity-violating process from modifications to the Higgs coupling to $W/Z$ bosons.  The left-hand side amplitudes are model-independent since they only depend on $\de_{V1}$ while the ones on right-hand side depend on $\de_{V2}$ as well.}
\label{tab:a}
\end{minipage}}
\end{table}

\begin{table}[!ht]
\centerline{\begin{minipage}{0.8\textwidth}
\centering
\vspace{1 mm}
\tabcolsep7pt\begin{tabular}{|c|c||c|c|}
\hline
 \textbf{Process} & \textbf{$\times\frac{(\de_{V2}-4\de_{V1})E^3}{96\pi^2 v^3}$} & \textbf{Process} & \textbf{$\times\frac{(\de_{V2}-4\de_{V1})E^3}{96\pi^2 v^3}$} \\
\hline
$hW^{+}W^{+} \rightarrow W^{+}W^{+}$ & $\sqrt{2}$ & $hW^{-}W^{+} \rightarrow ZZ$ & $-2$\\
$hW^{+}W^{-} \rightarrow W^{+}W^{-}$ & $-\sqrt{2}$ & $ZW^{-}W^{+} \rightarrow hZ$ & $0$\\
$W^{-}W^{+}W^{+} \rightarrow hW^{+}$ & $0$ & $Z^3 \rightarrow hZ$ & $0$ \\
$ZZW^{+} \rightarrow hW^{+}$ & $0$ & $Z^2h \rightarrow Z^2$ & $0$ \\
$hZW^{+} \rightarrow ZW^{+}$ & $\sqrt{2}$ & $Z^2h \rightarrow W^{+}W^{-}$ & $-2$\\
\hline
\end{tabular}
\vspace{1 mm}
\caption{\small 5-body unitarity-violating processes that depend on $\de_{V2}$ and $\de_{V1}$. One can see that the dim-6 SMEFT prediction $\de_{V2} = 4\de_{V1}$ gives vanishing amplitudes for all processes.}
\label{tab:b}
\end{minipage}}
\end{table}

\begin{table}[!ht]
\centerline{\begin{minipage}{0.8\textwidth}
\centering
\vspace{1 mm}
\tabcolsep7pt\begin{tabular}{|c|c||c|c|}
\hline
 \textbf{Process} & \textbf{$\times \frac{\delta_{Z1}E^{2}}{8\pi v^{2}}$} &  \textbf{Process} & \textbf{$\times \frac{\delta_{Z1}E^{3}}{24 \pi^{2} v^{3}} $} \\
\hline
$ZZ \rightarrow ZZ$ & $0$ & $W^{+}W^{-} \rightarrow Z^{3}$ & $0$\\
$ZZ \rightarrow W^{+}W^{-}$ & $-\frac{1}{\sqrt{2}} \Big(1+\lambda_{WZ}\Big) $& $ZW^{+} \rightarrow Z^{2}W^{+} $& $0$\\ 
$ZW^{+} \rightarrow ZW^{+}$ & $\frac{1}{2} \Big(1+\lambda_{WZ} \Big) $& $Z^{2} \rightarrow ZW^{+}W^{-} $& $0$\\ 
$W^{+}W^{-} \rightarrow W^{+}W^{-}$ & $-\lambda_{WZ}$& $W^{+}W^{-} \rightarrow ZW^{+}W^{-} $& $0$\\ 
$W^{+}W^{+} \rightarrow W^{+}W^{+}$ & $\lambda_{WZ}$& $W^{+}W^{+} \rightarrow ZW^{+}W^{+} $& $0$\\ 
$hW^{+} \rightarrow ZW^{+}$ & $\frac{3i}{2} \Big(1-\lambda_{WZ} \Big)$& $ZW^{+} \rightarrow W^{+}W^{-}W^{+}$& $i\Big( 1-\lambda_{WZ}\Big)$\\ 
$W^{+}W^{-} \rightarrow h Z$ & 0 & & \\ 
\hline
\end{tabular}
\vspace{1 mm}
\caption{\small 4-body and some 5-body unitarity-violating processes without assuming custodial symmetry. Here $\lambda_{WZ} = \frac{\delta_{W1}}{\delta_{Z1}} = 1$ in the custodial-preserving limit.}
\label{tab:c}
\end{minipage}}
\end{table}

\begin{table}[!ht]
\centerline{\begin{minipage}{0.8\textwidth}
\centering
\vspace{1 mm}
\tabcolsep7pt\begin{tabular}{|c|c|}
\hline
 \textbf{Process} & \textbf{$\times \frac{\left(\de_{V2}-4\de_{V1}\right)E^4}{384\pi^3 v^4}$} \\
\hline
$ZZZ \rightarrow ZZZ$ & 0 \\ [6pt]
$W^{+}W^{+}W^{+} \rightarrow W^{+}W^{+}W^{+}$ & $1$ \\ [6pt]
$ZW^{+}W^{+} \rightarrow ZW^{+}W^{+}$ & $1$ \\ [6pt]
$ZW^{+}W^{-} \rightarrow ZZZ$ & $-\sqrt{\frac{2}{3}}$ \\ [6pt]
$ZZW^{+} \rightarrow W^{+}W^{+}W^{-}$ & $-\frac{2}{3}$  \\ [6pt]
$ZZW^{+} \rightarrow ZZW^{+}$ & $\frac{2}{3}$ \\ [6pt]
$ZW^{+}W^{-} \rightarrow ZW^{+}W^{-}$ & $\frac{1}{3}$ \\ [6pt]
$W^{+}W^{+}W^{-} \rightarrow W^{+}W^{+}W^{-}$ & $-\frac{1}{3}$ \\ [6pt]
\hline
\end{tabular}
\vspace{1 mm}
\caption{\small 6-body unitarity-violating processes that depend on $\de_{V2}$ and $\de_{V1}$. One can see that the dim-6 SMEFT prediction $\de_{V2} = 4\de_{V1}$ gives vanishing amplitudes for all processes.}
\label{tab:d}
\end{minipage}}
\end{table}

\begin{table}[!ht]
\centerline{\begin{minipage}{0.8\textwidth}
\centering
\vspace{1 mm}
\tabcolsep7pt\begin{tabular}{|c|c|}
\hline
\textbf{Process} & \textbf{$ \times \frac{E^{4}}{1152 \pi^{3} v^{4}}$}  \\ [6pt]
\hline
 $h Z^{2} \rightarrow h Z^{2} $ & $[4 \de_{V1} - 2 \de_{V2} + \frac{1}{2} c_{V3} ]	 $ \\[6pt]
$h^{2} Z \rightarrow Z^{3} $ & $ -\frac{\sqrt{3}}{2}[4 \de_{V1} - 2 \de_{V2} + \frac{1}{2} c_{V3} ]$ \\ [6pt]
$h^{2} W^{+} \rightarrow Z^{2} W^{+} $ & $ -\frac{1}{2}[4 \de_{V1} - 2 \de_{V2} + \frac{1}{2} c_{V3} ]$ \\[6pt]
$h^{2} Z \rightarrow Z W^{+} W^{-} $ & $ -\frac{1}{\sqrt{2}}[4 \de_{V1}- 2 \de_{V2} + \frac{1}{2} c_{V3} ]$\\  [6pt]
$h^{2} W^{+} \rightarrow W^{+} W^{-} W^{+} $ & $ - [4 \de_{V1} - 2 \de_{V2} + \frac{1}{2} c_{V3} ]$ \\[6pt]
$h Z W^{+} \rightarrow  h Z W^{+} $ & $  [36 \de_{V1} - 13 \de_{V2} + 2 c_{Vc} ]$\\  [6pt]
$h W^{+} W^{+} \rightarrow  h W^{+} W^{+} $ & $ [36 \de_{V1} - 13 \de_{V2} + 2 c_{V3} ] $ \\[6pt]
 $h W^{+} W^{-} \rightarrow  h W^{+} W^{-} $ & $ - [28 \de_{V1} - 9 \de_{V2} + c_{V3} ]$\\ [6pt]
 $h Z^{2}\rightarrow  h W^{+} W^{-} $ & $ -\sqrt{2} [32 \de_{V1} - 11 \de_{V2} + \frac{3}{2} c_{V3} ]$\\ [6pt]
\hline
\end{tabular}
\vspace{1 mm}
\caption{\small 6-body  unitarity-violating processes that depend on $\de_{V1}$, $\de_{V2}$, and $c_{V3}$. One can see that the dim-6 SMEFT prediction $\de_{V2} = 4\de_{V1}$ and $c_{V3} = 8 \de_{V1}$ gives vanishing amplitudes for all processes.}
\label{tab:e}
\end{minipage}}
\end{table}

\begin{table}[!ht]
\centerline{\begin{minipage}{0.8\textwidth}
\centering
\vspace{1 mm}
\tabcolsep7pt\begin{tabular}{|c|c||c|c|}
\hline
\textbf{Process} & \textbf{$\times\frac{m_{t}\de_{t1}E}{8\pi v^2}$} &  \textbf{Process} & \textbf{$\times\frac{m_{t}\de_{t1}E}{8\pi v^2}$} \\ [6pt]
\hline
$\widebar{t}_{R}t_{R} \rightarrow Zh$ & $i\sqrt{N_c}$ & $t_{R}W^{+} \rightarrow t_{L}W^{+}$ & $-\frac{1}{2}$\\ [6pt]
$\widebar{t}_{R}t_{R} \rightarrow ZZ$ & $-\sqrt{\frac{N_c}{2}}$ & $\widebar{b}_{R}t_{R} \rightarrow hW^{+}$ & $\sqrt{2N_c}$ \\ [6pt]
$\widebar{t}_{R}t_{R} \rightarrow W^{-}W^{+}$ & $-\sqrt{N_c}$ & $t_{R}h \rightarrow b_{L}W^{+}$ & $\frac{1}{\sqrt{2}}$ \\ [6pt]
$t_{R}Z \rightarrow t_{L}h$ & $\frac{i}{2}$ & $t_{R}W^{-} \rightarrow b_{L}h$ & $\frac{1}{\sqrt{2}}$ \\ [6pt]
$t_{R}Z \rightarrow t_{L}Z$ & $-\frac{1}{2}$ &  & \\ [6pt]
\hline
\textbf{Process} & $\times\frac{m_{t}c_{t2}E}{8\pi v^2}$ &  \textbf{Process} & \textbf{$\times\frac{m_{t}c_{t2}E}{8\pi v^2}$} \\ [6pt]
\hline
$\bar{t}_{R}t_{R} \rightarrow hh$ &$-\sqrt{\frac{N_{c}}{2}} $ &  $t_{R} h \rightarrow t_{L} h$ & $-\frac{1}{2}$ \\ [6pt]
\hline
\end{tabular}
\vspace{1 mm}
\caption{\small 4-body model-independent unitarity-violating processes from the top sector.}
\label{tab:f}
\end{minipage}}
\end{table} 

\begin{table}[!ht]
\centerline{\begin{minipage}{0.8\textwidth}
\centering
\vspace{1 mm}
\tabcolsep7pt\begin{tabular}{|c|c||c|c|}
\hline
\textbf{Process} & \textbf{$\times\frac{m_{t}\de_{t1}E^2}{64\pi^2 v^3}$} & \textbf{Process} & \textbf{$\times\frac{m_{t}\de_{t1}E^2}{64\pi^2 v^3}$} \\ [6pt]
\hline
$\widebar{t}_{R}t_{R} \rightarrow ZZZ$ & $i\sqrt{3N_c}$ & $Z^2 \rightarrow \widebar{t}_{L}b_{L}W^{+}$ & $\sqrt{\frac{2N_c}{3}}$\\ [6pt]
$\widebar{t}_{R}t_{R} \rightarrow ZW^{+}W^{-}$ & $i\sqrt{2N_c}$ & $ZW^{-} \rightarrow Zb_{L}\widebar{t}_{L}$ & $2\sqrt{\frac{N_c}{3}}$\\ [6pt]
$t_{R}Z \rightarrow t_{L}W^{-}W^{+}$ & $\frac{i}{\sqrt{3}}$ & $t_{R}Z \rightarrow b_{L}ZW^{+}$ & $\sqrt{\frac{2}{3}}$\\ [6pt]
$t_{R}Z \rightarrow t_{L}ZZ$ & $i\sqrt{\frac{3}{2}}$ & $t_{R}W^{-} \rightarrow b_{L}Z^2$ & $\frac{1}{\sqrt{3}}$\\ [6pt]
$t_{R}W^{+} \rightarrow t_{L}ZW^{+}$ & $\frac{i}{\sqrt{3}}$ & $\widebar{b}_{R}t_{R} \rightarrow W^{+}W^{+}W^{-}$ & $2\sqrt{2N_c}$\\ [6pt]
$W^{+}W^{-} \rightarrow \widebar{t}_{L}t_{L}Z$ & $i\sqrt{\frac{2N_c}{3}}$ & $W^{-}W^{-} \rightarrow b_{L}\widebar{t}_{L}W^{-}$ & $2\sqrt{\frac{2N_c}{3}}$\\ [6pt]
$W^{+}Z \rightarrow \widebar{t}_{L}t_{L}W^{+}$ & $i\sqrt{\frac{2N_c}{3}}$ & $W^{+}W^{-} \rightarrow b_{L}\widebar{t}_{L}W^{+}$ & $4\sqrt{\frac{N_c}{3}}$\\ [6pt]
$ZZ \rightarrow \widebar{t}_{L}t_{L}Z$ & $i\sqrt{3N_c}$ & $t_{R}W^{+} \rightarrow b_{L}W^{+}W^{+}$ & $2\sqrt{\frac{N_c}{3}}$\\ [6pt]
$\widebar{b}_{R}t_{R} \rightarrow Z^2W^{+}$ & $\sqrt{2N_c}$ & $t_{R}W^{-} \rightarrow b_{L}W^{-}W^{+}$ & $2\sqrt{\frac{2N_c}{3}}$\\ [6pt]
\hline
\end{tabular}
\vspace{1 mm}
\caption{\small 5-body model-independent unitarity-violating processes from the top sector.}
\label{tab:g}
\end{minipage}}
\end{table} 

\begin{table}[!ht]
\centerline{\begin{minipage}{0.8\textwidth}
\centering
\vspace{1 mm}
\tabcolsep7pt\begin{tabular}{|c|c||c|c|}
\hline
 \textbf{Process} & \textbf{$\times \frac{ \left(\frac{1}{2}c_{t2}- \de_{t1} \right)m_{t} E^2}{32\pi^2 v^3}$} &  \textbf{Process} & \textbf{$\times \frac{ \left(\frac{1}{2}c_{t2}- \de_{t1} \right)m_{t} E^2}{32\pi^2 v^3}$} \\ [6pt]
\hline
$\widebar{t}_{R}t_{R} \rightarrow Zh^2$ & $i\sqrt{N_c}$ & $\widebar{t}_{R}t_{R} \rightarrow W^{+}W^{-}h$ & $-\sqrt{2N_c}$\\ [6pt]
$h^2 \rightarrow Z\widebar{t}_{L}t_{L}$ & $i\sqrt{\frac{N_c}{3}}$ & $W^{+}W^{-} \rightarrow \widebar{t}_{L}t_{L}h$ & $-\sqrt{\frac{2N_c}{3}}$\\ [6pt]
$Zh \rightarrow h\widebar{t}_{L}t_{L}$ & $i\sqrt{\frac{2N_c}{3}}$ & $W^{+}h \rightarrow \widebar{t}_{L}t_{L}W^{+}$ & $-\sqrt{\frac{2N_c}{3}}$\\ [6pt]
$t_{R}Z \rightarrow t_{L}h^2$ & $\frac{i}{\sqrt{6}}$ & $t_{R}W^{+} \rightarrow t_{L}W^{+}h$ & $-\frac{1}{\sqrt{3}}$\\ [6pt]
$t_{R}h \rightarrow t_{L}Zh$ & $\frac{i}{\sqrt{3}}$ & $t_{R}h \rightarrow t_{L}W^{+}W^{-}$ & $-\frac{1}{\sqrt{3}}$\\ [6pt]
$\widebar{t}_{R}t_{R} \rightarrow Z^2h$ & $-\sqrt{N_c}$ & $\widebar{b}_{R}t_{R} \rightarrow W^{+}h^2$ & $\sqrt{2N_c}$\\ [6pt]
$Z^2 \rightarrow \widebar{t}_{L}t_{L}h$ & $-\sqrt{\frac{N_c}{3}}$ & $W^{-}h \rightarrow b_{L}\widebar{t}_{L}h$ & $2\sqrt{\frac{N_c}{3}}$\\ [6pt]
$Zh \rightarrow \widebar{t}_{L}t_{L}Z$ & $-\sqrt{\frac{2N_c}{3}}$ & $h^2 \rightarrow b_{L}\widebar{t}_{L}W^{+}$ & $\sqrt{\frac{2N_c}{3}}$\\ [6pt]
$t_{R}h \rightarrow t_{L}Z^2$ & $-\frac{1}{\sqrt{6}}$ & $t_{R}W^{-} \rightarrow b_{L}h^2$ & $\frac{1}{\sqrt{3}}$\\ [6pt]
$t_{R}Z \rightarrow t_{L}Zh$ & $-\frac{1}{\sqrt{3}}$ & $t_{R}h \rightarrow b_{L}W^{+}h$ & $\sqrt{\frac{2}{3}}$\\ [6pt]
\hline
\end{tabular}
\vspace{1 mm}
\caption{\small 5-body unitarity-violating processes that depend on $c_{t2}$ and $\de_{t1}$.}
\label{tab:h}
\end{minipage}}
\end{table} 

\begin{table}[!ht]
\centerline{\begin{minipage}{0.8\textwidth}
\centering
\vspace{1 mm}
\tabcolsep7pt\begin{tabular}{|c|c||c|c|}
\hline
 \textbf{Process} & \textbf{$\times \frac{\left(3\de_{t1}-c_{t2}\right)m_{t}E^3}{256\pi^3v^4}$} & \textbf{Process} & \textbf{$\times \frac{\left(3\de_{t1}-c_{t2}\right)m_{t}E^3}{256\pi^3v^4}$} \\ [6pt]
\hline
$\widebar{t}_{R}t_{R}Z \rightarrow Z^3$ & $\sqrt{\frac{N_c}{2}}$ & $t_{R}Z^2 \rightarrow t_{L}Zh$ & $-\frac{i}{\sqrt{2}}$\\ [6pt]
$t_{R}Z^2 \rightarrow t_{L}Z^2$ & $\frac{1}{2}$ & $\widebar{t}_{R}t_{R}Z \rightarrow hW^{+}W^{-}$ & $-i\sqrt{\frac{N_c}{3}}$\\ [6pt]
$\widebar{t}_{R}t_{R}W^{+} \rightarrow Z^2W^{+}$ & $\sqrt{\frac{N_c}{6}}$ & $t_{R}Zh \rightarrow t_{L}W^{+}W^{-}$ & $-\frac{i}{3}$\\ [6pt]
$\widebar{t}_{R}t_{R}Z \rightarrow ZW^{+}W^{-}$ & $\sqrt{\frac{N_c}{3}}$ & $\widebar{b}_{R}t_{R}W^{-} \rightarrow hZ^2$ & $-\sqrt{\frac{N_c}{3}}$\\ [6pt]
$t_{R}Z^2 \rightarrow t_{L}W^{+}W^{-}$ & $\frac{1}{3\sqrt{2}}$ & $\widebar{b}_{R}t_{R}Z \rightarrow hZW^{+}$ & $-\sqrt{\frac{2N_c}{3}}$\\ [6pt]
$t_{R}ZW^{+} \rightarrow t_{L}ZW^{+}$ & $\frac{1}{3}$ & $t_{R}Z^2 \rightarrow b_{L}W^{+}h$ & $-\frac{1}{3}$\\ [6pt]
$\widebar{t}_{R}t_{R}W^{+} \rightarrow W^{+}W^{+}W^{-}$ & $\sqrt{\frac{2N_c}{3}}$ & $t_{R}Zh \rightarrow b_{L}ZW^{+}$ & $-\frac{\sqrt{2}}{3}$\\ [6pt]
$t_{R}W^{+}W^{+} \rightarrow t_{L}W^{+}W^{+}$ & $\frac{1}{3}$ & $\widebar{b}_{R}t_{R}h \rightarrow W^{+}W^{+}W^{-}$ & $-2\sqrt{\frac{N_c}{3}}$\\ [6pt]
$t_{R}W^{+}W^{-} \rightarrow t_{L}W^{+}W^{-}$ & $\frac{2}{3}$ & $\widebar{b}_{R}t_{R}W^{-} \rightarrow hW^{+}W^{-}$ & $-2\sqrt{\frac{2N_c}{3}}$\\ [6pt]
$\widebar{t}_{R}t_{R}h \rightarrow Z^3$ & $-i\sqrt{\frac{N_c}{2}}$ & $t_{R}W^{-}W^{-} \rightarrow b_{L}W^{-}h$ & $-\frac{2}{3}$\\ [6pt]
$\widebar{t}_{R}t_{R}Z \rightarrow Z^2h$ & $-i\sqrt{\frac{3N_c}{2}}$ & $t_{R}W^{-}h \rightarrow b_{L}W^{+}W^{-}$ & $-\frac{2\sqrt{2}}{3}$\\ [6pt]
$ \bar{b}_{R}t_{R}W^{+} \rightarrow hW^{+}W^{+}$ & $-2\sqrt{\frac{N_{c}}{3}}$ &  & \\ [6pt]
\hline
\end{tabular}
\vspace{1 mm}
\caption{\small 6-body unitarity-violating processes that depend on $c_{t2}$ and $\de_{t1}$. One can see that the dim-6 SMEFT prediction $c_{t2} = 3\de_{t1}$ gives vanishing amplitudes for all processes.}
\label{tab:i}
\end{minipage}}
\end{table} 

\begin{table}[!ht]
\centerline{\begin{minipage}{0.8\textwidth}
\centering
\vspace{1 mm}
\end{minipage}}
\tabcolsep7pt\begin{tabular}{|c|c||c|c|}
\hline
 \textbf{Process} & \textbf{$\times\frac{\left(c_{t2}-3\de_{t1} \right)m_{t} E^4}{1024\pi^4 v^5}$} &  \textbf{Process} & \textbf{$\times\frac{\left(c_{t2}-3\de_{t1} \right)m_{t} E^4}{1024\pi^4 v^5}$} \\ [6pt]
\hline
$\widebar{t}_{R}t_{R}Z \rightarrow Z^4$ & $\frac{5i}{4}\sqrt{\frac{N_c}{3}}$ & $ZW^{+}W^{+} \rightarrow \widebar{t}_{L}t_{L}W^{+}W^{+}$ & $\frac{i\sqrt{N_c}}{6}$ \\ [6pt]
$\widebar{t}_{R}t_{R}W^{+} \rightarrow W^{+}Z^3$ & $\frac{i}{2}\sqrt{\frac{N_c}{3}}$ & $ZW^{+}W^{-} \rightarrow \widebar{t}_{L}t_{L}Z^2$ & $\frac{i}{2}\sqrt{\frac{N_c}{2}}$\\ [6pt]
$\widebar{t}_{R}t_{R}W^{+} \rightarrow ZW^{-}W^{+}W^{+}$ & $\frac{i\sqrt{N_c}}{3}$ & $ZW^{+}W^{-} \rightarrow \widebar{t}_{L}t_{L}W^{+}W^{-}$ & $\frac{i\sqrt{N_c}}{3}$\\ [6pt]
$\widebar{t}_{R}t_{R}Z \rightarrow W^{+}W^{+}W^{-}W^{-}$ & $\frac{i}{3}\sqrt{\frac{N_c}{2}}$ & $W^{+}W^{+}W^{-} \rightarrow \widebar{t}_{L}t_{L}ZW^{+}$ & $\frac{i}{3}\sqrt{\frac{N_c}{2}}$\\ [6pt]
$t_{R}Z^2 \rightarrow t_{L}ZW^{-}W^{+}$ & $\frac{i}{4}$ & $\widebar{b}_{R}t_{R}W^{-} \rightarrow Z^4$ & $\frac{1}{4}\sqrt{\frac{2N_c}{3}}$\\ [6pt]
$t_{R}Z^2 \rightarrow t_{L}Z^3$ & $\frac{5i}{4\sqrt{6}}$ & $\widebar{b}_{R}t_{R}Z \rightarrow W^{+}Z^3$ & $\frac{1}{2}\sqrt{\frac{2N_c}{3}}$\\ [6pt]
$t_{R}W^{+}W^{+} \rightarrow t_{L}ZW^{+}W^{+}$ & $\frac{i}{6\sqrt{2}}$ & $Z^3 \rightarrow b_{L}\widebar{t}_{L}ZW^{+}$ & $\frac{1}{2}\sqrt{\frac{N_c}{3}}$\\ [6pt]
$t_{R}W^{-}W^{+} \rightarrow t_{L}Z^3$ & $\frac{i}{4\sqrt{3}}$ & $Z^2W^{-} \rightarrow b_{L}\widebar{t}_{L}Z^2$ & $\frac{1}{2}\sqrt{\frac{N_c}{2}}$\\ [6pt]
$t_{R}W^{-}W^{+} \rightarrow t_{L}ZW^{+}W^{-}$ & $\frac{i}{3\sqrt{2}}$ & $t_{R}Z^2 \rightarrow b_{L}W^{+}Z^2$ & $\frac{1}{4}$\\ [6pt]
$t_{R}ZW^{+} \rightarrow t_{L}Z^2W^{+}$ & $\frac{i}{4}$ & $t_{R}ZW^{-} \rightarrow b_{L}Z^3$ & $\frac{\sqrt{2}}{4\sqrt{3}}$\\ [6pt]
$t_{R}ZW^{+} \rightarrow t_{L}W^{+}W^{+}W^{-}$ & $\frac{i}{6}$ & $\widebar{b}_{R}t_{R}W^{+} \rightarrow W^{+}W^{+}Z^2$ & $\frac{\sqrt{N_c}}{3}$\\ [6pt]
$Z^3 \rightarrow \widebar{t}_{L}t_{L}Z^2$ & $\frac{5i}{4}\sqrt{\frac{N_c}{3}}$ & $\widebar{b}_{R}t_{R}W^{-} \rightarrow Z^2W^{+}W^{-}$ & $\frac{\sqrt{2N_c}}{3}$\\ [6pt]
$Z^3 \rightarrow \widebar{t}_{L}t_{L}W^{+}W^{-}$ & $\frac{i}{2}\sqrt{\frac{N_c}{6}}$ & $W^{+}W^{-}W^{-} \rightarrow b_{L}\widebar{t}_{L}Z^2$ & $\frac{1}{3}\sqrt{\frac{N_c}{2}}$\\ [6pt]
$Z^2W^{+} \rightarrow \widebar{t}_{L}t_{L}ZW^{+}$ & $\frac{i}{2}\sqrt{\frac{N_c}{2}}$ & $ZW^{-}W^{-} \rightarrow b_{L}\widebar{t}_{L}ZW^{-}$ & $\frac{\sqrt{N_c}}{3}$\\ [6pt]
$t_{R}W^{-}W^{-} \rightarrow b_{L}W^{-}Z^2$ & $\frac{1}{6}$ & $W^{+}W^{-}W^{-} \rightarrow b_{L}\widebar{t}_{L}W^{-}W^{+}$ & $\sqrt{N_c}$\\ [6pt]
$t_{R}W^{+}W^{-} \rightarrow b_{L}Z^2W^{+}$ & $\frac{\sqrt{2}}{6}$ & $W^{+}W^{+}W^{-} \rightarrow b_{L}\widebar{t}_{L}W^{+}W^{+}$ & $\sqrt{\frac{N_c}{2}}$\\ [6pt]
$t_{R}ZW^{-} \rightarrow b_{L}ZW^{-}W^{+}$ & $\frac{1}{3}$ & $t_{R}W^{+}W^{+} \rightarrow b_{L}W^{+}W^{+}W^{+}$ & $\frac{1}{2\sqrt{3}}$\\ [6pt]
$\widebar{b}_{R}t_{R}W^{+} \rightarrow W^{+}W^{+}W^{+}W^{-}$ & $\sqrt{\frac{2N_c}{3}}$ & $t_{R}W^{-}W^{-} \rightarrow b_{L}W^{+}W^{-}W^{-}$ & $\frac{1}{2}$\\ [6pt]
$\widebar{b}_{R}t_{R}W^{-} \rightarrow W^{-}W^{-}W^{+}W^{+}$ & $\sqrt{N_c}$ & $t_{R}W^{+}W^{-} \rightarrow b_{L}W^{-}W^{+}W^{+}$ & $\frac{1}{\sqrt{2}}$\\ [6pt]
$W^{-}W^{-}W^{-} \rightarrow b_{L}\widebar{t}_{L}W^{-}W^{-}$ & $\sqrt{\frac{N_c}{6}}$ & $ t_{R} Z^{2} \rightarrow b_{R} W^{+}W^{-}W^{+} $ & $\frac{1}{6}$ \\ [6pt]
$ \bar{t}_{R}t_{R} Z \rightarrow W^{+}W^{-}Z^{2}$ & $\frac{i\sqrt{N_{c}}}{2}$ & $ \bar{b}_{R}t_{R} Z W^{-} \rightarrow W^{+}W^{-}Z $ & $\frac{\sqrt{2N_{c}}}{3}$ \\ [6pt]
$ \bar{b}_{R}t_{R}W^{+}W^{-} \rightarrow Z^{2}W^{+}$ & $\frac{\sqrt{N_{c}}}{3}$ & $ \bar{b}_{R}t_{R}Z \rightarrow W^{+}W^{-}W^{+}Z $ & $\frac{\sqrt{2N_{c}}}{3}$ \\ [6pt]
\hline
\end{tabular}
\vspace{1 mm}
\caption{\small 7-body unitarity-violating processes that depend on $c_{t2}$ and $\de_{t1}$. One can see that the dim-6 SMEFT prediction $c_{t2} = 3\de_{t1}$ gives vanishing amplitudes for all processes.}
\label{tab:j}
\end{table}
\clearpage
\bibliographystyle{utphys}
\bibliography{Kappa_Unitarity}

\end{document}